\documentclass[	pra,
		twocolumn,
		groupedaddress,
		nofootinbib,
		showpacs,
		showkeys,
		eqsecnum,	
		]{revtex4-1}
\usepackage{amsmath,amstext,amsfonts,amssymb,amsthm}
\usepackage{ wasysym }
\usepackage{bm}
\usepackage[caption=false]{subfig} 
\usepackage{dsfont}					
\usepackage[dvipsnames]{xcolor}
\usepackage{pgf}
\usepackage{tikz}
\usetikzlibrary{arrows,automata}
\usepackage{xcolor}

\definecolor{pine}{RGB}{59,119,86}
\definecolor{lavendel}{RGB}{180,102,255}
\definecolor{gold}{RGB}{255,150,50}
\definecolor{azure}{RGB}{120,150,220}
\definecolor{turquoise}{RGB}{0,191,191}
\definecolor{mygenta}{RGB}{255,0,255}

\newcommand{\ket}[1]{\left| #1\right>}
\newcommand{\bra}[1]{\left< #1\right|}

\newcommand{\vek}[1]{\mathbf #1}
\newcommand{\ketbra}[2]{|#1\rangle\!\langle #2|}
\newcommand{\braket}[2]{\left\langle#1\vphantom{#2}\right|\left.\vphantom{#1}#2\right\rangle}
\newcommand{\matrixElement}[3]{\left \langle   #1 \vphantom{#3}  \right |   #2  \left | \vphantom{#1} #3 \right \rangle}
\newcommand{\matrixElementSub}[5]{\mspace{3.0mu}{}_{#2}\mspace{-5.0mu}\matrixElement{#1}{#3}{#4}\mspace{-5.0mu}{}_{#5}}
\newcommand{\ketbraIX}[3]{\ket{#1}_{\! #2}\!\!\bra{#3}}
\newcommand{\braketIX}[4]{{\vphantom \langle}_{#1}\!\left\langle #2 \!\right.\left| #3\right\rangle_{#4}}
\newcommand{\creation}[1]{{{#1}^{\dagger}}}
\newcommand{\annihilation}[1]{{{#1}^{\vphantom{\dagger}}}}

\newcommand{\modulus}[1]{\left \lvert #1 \right \rvert}

\newcommand{\matrixNorm}[1]{\lVert #1 \rVert }
\newcommand{\commutator}[2]{ \left[ #1 , #2 \right] }
\DeclareMathOperator{\trace}{Tr}

\DeclareMathOperator{\Real}{Re}

\DeclareMathOperator{\diag}{diag}
\DeclareMathOperator{\atanh}{arctanh}
\DeclareMathAlphabet{\mathcalligra}{T1}{calligra}{m}{n}
\DeclareMathAlphabet{\mathpzc}{OT1}{pzc}{m}{it}

\begin{document}
\title{Quantum quenches of ion Coulomb crystals across structural instabilities}
\author{Jens D. \surname{Baltrusch}$^{1,2}$}
\email[Email: ]{jens.baltrusch@physik.uni-saarland.de}
\author{Cecilia \surname{Cormick}$^{1}$}
\author{Giovanna \surname{Morigi}$^{1}$}
 \affiliation{
$^1$ Theoretische Physik, Universit\"at des Saarlandes, D-66123 Saarbr\"ucken, Germany\\
$^2$Grup d'\`Optica, Departament de F\'isica, Universitat Aut\`onoma de Barcelona, E-08193 Bellaterra, Spain 
}
\date{\today
}
\begin{abstract}
Quenches in an ion chain can create coherent superpositions of motional states across the linear-zigzag structural transition. The procedure has been described in [Phys. Rev. A {\bf 84}, 063821 (2011)] and makes use of spin-dependent forces, so that a coherent superposition of the electronic states of one ion evolves into an entangled state between the chain's internal and external degrees of freedom. The properties of the crystalline state so generated are theoretically studied by means of Ramsey interferometry on one ion of the chain. An analytical expression for the visibility of the interferometric measurement is obtained for a chain of arbitrary number of ions and as a function of the time elapsed after the quench. Sufficiently close to the linear-zigzag instability the visibility decays very fast, but exhibits revivals at the period of oscillation of the mode that drives the structural instability. These revivals have a periodicity that is independent of the crystal size, and they signal the creation of entanglement by the quantum quench.
\end{abstract}
\maketitle
\section{Introduction}
  \label{sec:introduction}
  
Quenches of quantum many-body systems provide important information on the thermodynamic properties of physical objects close to phase transitions. They give insight into the statistical mechanics of closed systems, and can find applications for quantum information~\cite{Polkovnikov,Cardy,Rieger}. Among several proposals discussed in the literature, some set-ups make use of the coupling with a quantum system, a spin, to drive a quantum phase transition in a larger physical object acting as environment. These dynamics are associated with decay of spin coherence that has been shown to exhibit universal features~\cite{Zanardi,Paz,Cormick,Rossini}. 

\begin{figure}[bthp]
  \centering
 \includegraphics[width=0.36\textwidth]{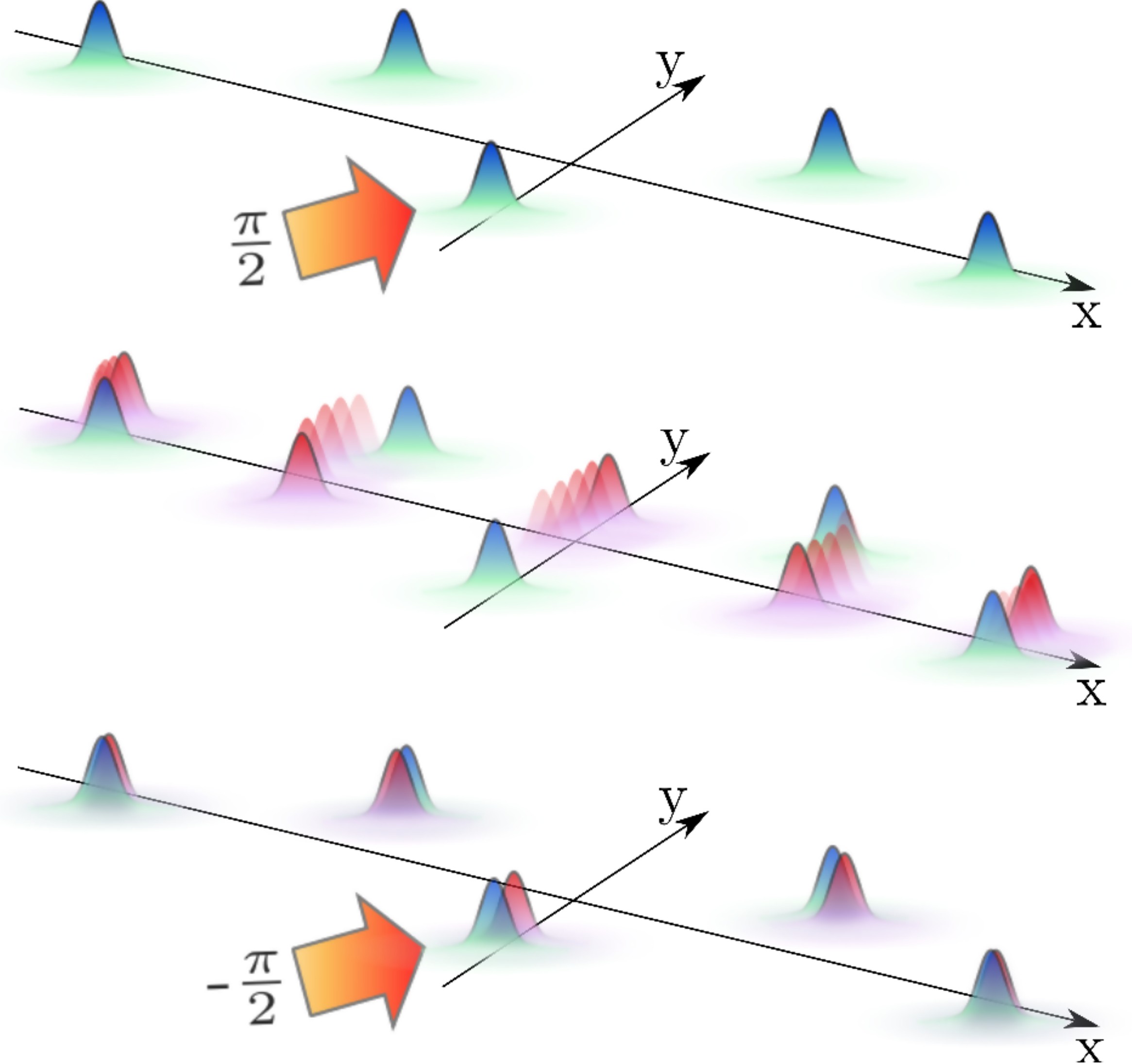} \put(-200,170){(a)} \put(-200,110){(b)}\put(-200,50){(c)}
 \caption{\label{Fig:1}(Color online) A quench across the linear-zigzag instability is performed by exciting the central ion with a laser pulse in presence of spin-dependent forces. In (a) the collective motion is initially in the ground state of a zigzag structure, and the central ion in the internal state $|g\rangle$. A $\pi/2$ laser pulse prepares it in the superposition $ (|g\rangle + |e\rangle)/\sqrt{2}$. (b) A tighter state-dependent
potential, acting only when the ion is in state $ |e\rangle$, induces conditional dynamics such that the ions' internal and external degrees of freedom get entangled. When the state-dependent potential is sufficiently tight, the excited component will start oscillating around the equilibrium positions of the linear chain. (c) A laser pulse performs a $-\pi/2$ rotation on the central ion. The final occupation of the ground state $|g\rangle$ as a function of the time between the two pulses contains information on the properties of the chain across the linear-zigzag instability.}
\end{figure}

In a recent article~\cite{DeChiara2008} these dynamics were studied in a system of trapped ions, when these form a linear array in a linear Paul trap\cite{Birkl1992}.
Here, the spin is an internal transition of one ion of the chain, while the vibrational excitations of the chain itself acts as an environment. Realizing a Ramsey-type of interferometer with the internal transition of the ion, a quench is performed by the mechanical effect of light associated with the absorption and the emission of the laser photon. This quench is carried out when the chain is close to the mechanical instability at which it undergoes a transition to a zigzag structure~\cite{Fishman2008}. It was shown that the visibility of the Ramsey interferometer, which is found by measuring the population of one internal state of the ion transition after the second Ramsey pulse, allows one to access the autocorrelation function of the chain at criticality~\cite{DeChiara2008}. 

In Ref.~\cite{Baltrusch2011}  it was proposed to use the spin excitation to create a superposition of two different crystalline structures across the linear-zigzag structural transition. The superposition can be accessed by driving the electronic transition of one ion of the chain in a set-up where an external field makes the trap frequency spin-dependent~\cite{Baltrusch2011, Li-Lesanovsky}.  In these settings, a first laser pulse prepares the ion in a coherent superposition of the electronic states, which evolves into an entangled state between the chain's internal and external degrees of freedom  as sketched in Fig.~\ref{Fig:1}. The properties of the crystalline state so generated were studied by evaluating numerically the visibility after a second laser pulse is applied, as shown in Fig.~\ref{Fig:1}(c).  The visibility of the interferometric signal was shown to exhibit a fast decay, in agreement with studies performed in other settings~\cite{Paz,Cormick,Rossini}, while for longer times quasi-periodic revivals of the visibility were found. 

In this paper we analyse the dependence of the signal visibility on the system parameters for the set-up proposed in Ref.~\cite{Baltrusch2011}. We determine analytically the expression of the visibility and study its behaviour close to and across the classical linear-zigzag instability, for different numbers of ions. We find that the revivals observed in the visibility as a function of the time $t$ elapsed after the quench are characterized by the frequency of the zigzag mode, and persist when the number of ions is increased. The analysis of the spectrum of the visibility signal as a function of $t$ shows the presence of squeezing and entanglement that are generated by the quantum quench. 

The article is organized as follows: In Sec.~\ref{sec:ramsey} the proposal of Ref.~\cite{Baltrusch2011} is summarized. The theoretical model is  presented in Sec.~\ref{sec:harmonic_model}, which also includes the detailed evaluation of the visibility signal. The behaviour of the visibility is analysed in  Sec.~\ref{sec:results}, and the conclusions are drawn in Sec.~\ref{sec:conclusions}. Theoretical details for the derivation of the results in Sec.~\ref{sec:harmonic_model} are given in the appendices.

\section{Ramsey Interferometry with an ion Coulomb crystal}
  \label{sec:ramsey}

In this section we briefly review the physical model at the basis of this work. A string of $N$ ions of mass $m$ and charge $q$ is confined in a trap, forming a zigzag structure close to the linear-zigzag mechanical instability. A two-level transition of the central ion is driven by two laser pulses separated by a time interval $t$ and which perform a $\pi/2$ and $-\pi/2$ rotation of the dipole, respectively. The pulses are short so that the crystal dynamics can be neglected during their duration~\cite{DeChiara2008, MonroePRL2010}. Under the assumption that both internal states of the dipolar transition are stable, a spin-dependent force is applied, such that the stable configuration of a finite chain is a linear structure when the ion is in the excited state~\cite{Baltrusch2011}. Therefore, during the time elapsed between the two pulses, the crystalline state undergoes conditional dynamics dependent on the internal state, that lead to entanglement between internal and external degrees of freedom~\cite{DeChiara2008,Baltrusch2011}. 

In the following we denote by $|g\rangle$ and $|e\rangle$ the two internal states of the central ion, and omit to write the internal state of the other $N-1$ ions since this remains unchanged. Before the first pulse, the state of the central ion and crystal motion reads  $\ket{\psi(0)} =|g\rangle \ket{\phi(0)}$ where $\ket{\phi(0)}$ can be either the ground state of the linear or of the zigzag configuration, as shown in Fig.~\ref{Fig:1}(a). 
The pulse performs a quantum quench by bringing the central ion into a superposition of ground and excited states. In fact, at a time $t$ after the first pulse the state takes the form:
\begin{equation}
 \label{eq:ramsey:state1}
\ket{\psi(t)} = \frac{1}{\sqrt{2}}\Big( \ket{g} \ket{\phi_{g}(t)} + {\rm e}^{{\rm i}\varphi}\ket{e} \ket{\phi_{e}(t)} \Big) \,,
\end{equation}
where $\varphi$ is a controllable phase and
\begin{equation}
\ket{\phi_{s}(t)}=\exp\left(-{\rm i}H_{s}t/\hbar\right)\ket{\phi(0)} \,,
\end{equation}
with $s=g,e$ and $H_g$, $H_e$ the Hamiltonians for the external degrees of freedom, accounting for the state-dependent potential. The free evolution is pictorially shown in Fig.~\ref{Fig:1}(b) and leads to entanglement between internal and external degrees of freedom. After the second pulse, which performs a  $-\pi/2$ rotation of the dipole as sketched in Fig.~\ref{Fig:1}(c), the probability of measuring the central ion in state $|g\rangle$ reads
\begin{equation} \label{eq:ramsey:Pg}
\mathcal{P}_g(\phi) = \frac{1}{2} \Big\{1 + \Real \left[ e^{i\varphi} \mathcal O(t)  \right] \Big\} \,,
\end{equation}
where 
\begin{equation}
  \label{eq:overlap}
 \mathcal O(t) = \langle\phi_{g}(t)|\phi_{e}(t)\rangle \,
\end{equation}
is the overlap between the two motional states. The contrast of the Ramsey fringes is given by 
\begin{equation}
  \label{eq:overlapmodulus}
\mathcal V(t)= \left| \mathcal O(t) \right| \,, 
\end{equation}
and depends on the time $t$ elapsed between the two Ramsey pulses. We note that the visibility of the interference signal is directly related to the Loschmidt echo, frequently used to describe the loss of coherence as a consequence of the interaction between a system and its environment~\cite{Cucchietti}. 

\begin{figure}[tbp]
  \centering
  \subfloat{\label{fig:signal_lin}\phantom{(a)}\includegraphics[width=0.35\textwidth]{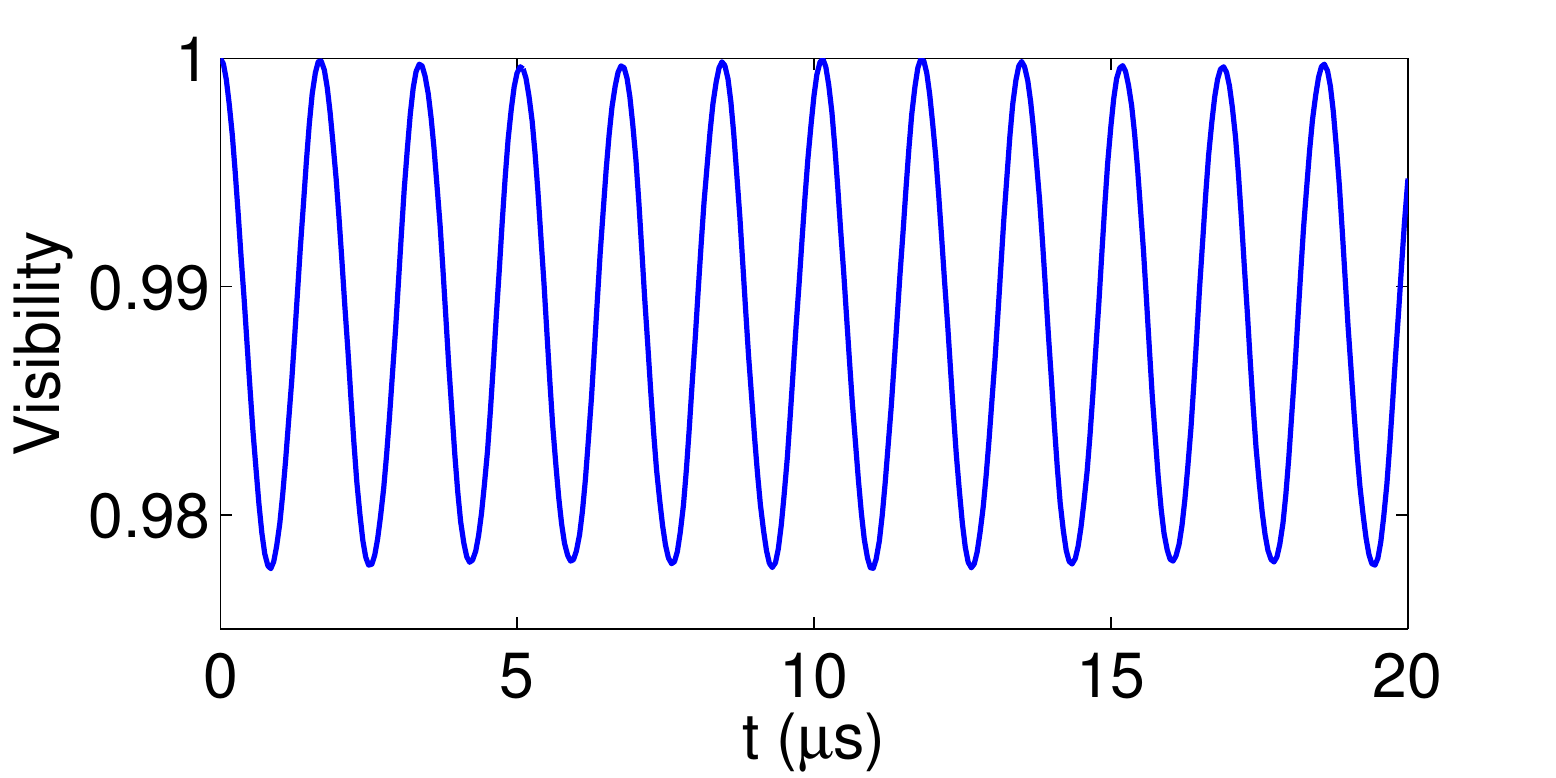}\put(-200,75){(a)}} 
  \\ \vspace{-.35cm}
  \subfloat{\label{fig:signal_zz}\phantom{(a)}\includegraphics[width=0.35\textwidth]{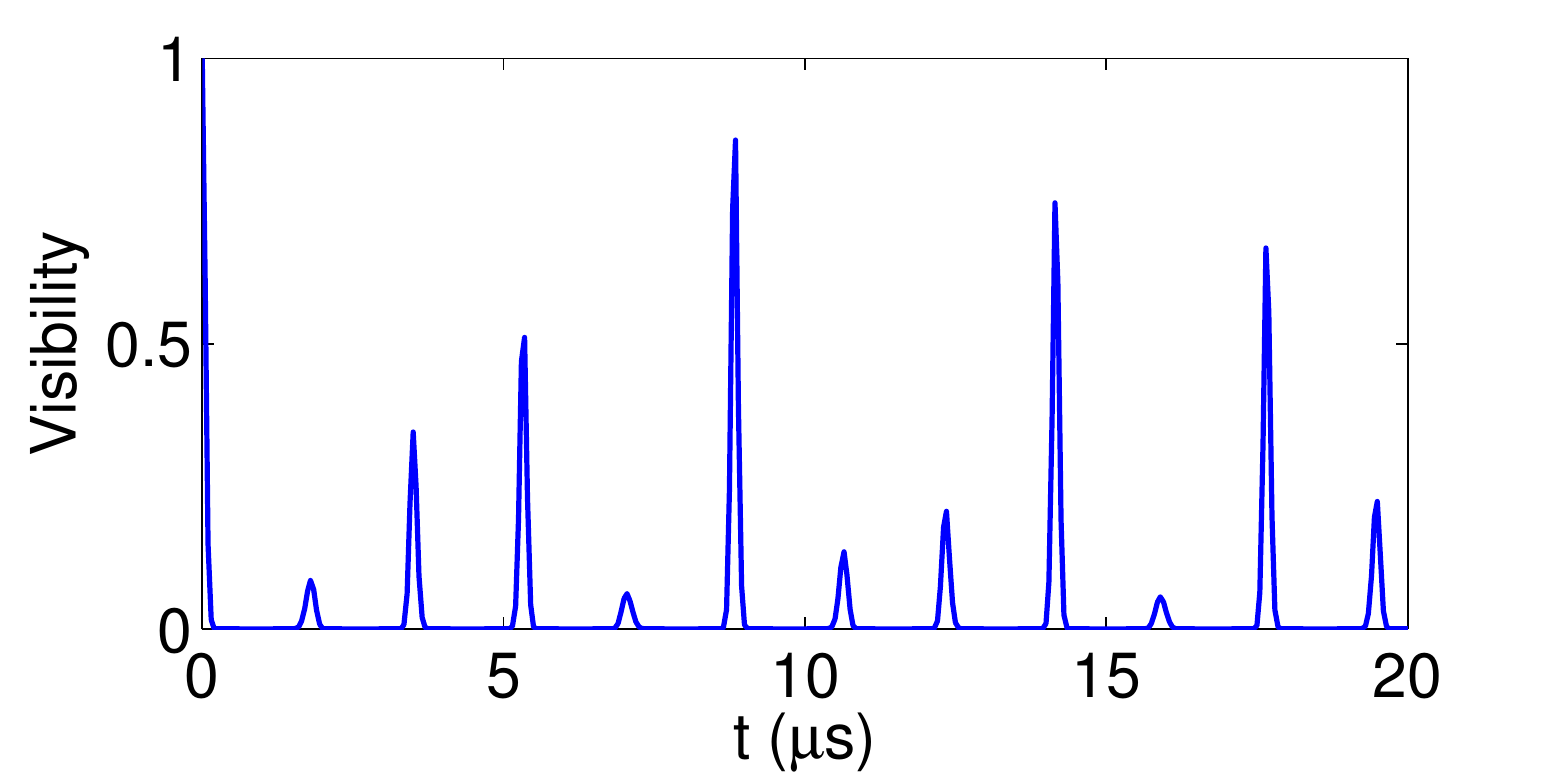}\put(-200,75){(b)}} \\ \vspace{-.35cm} 
  \subfloat{\label{fig:signal_sup}\phantom{(a)}\includegraphics[width=0.35\textwidth]{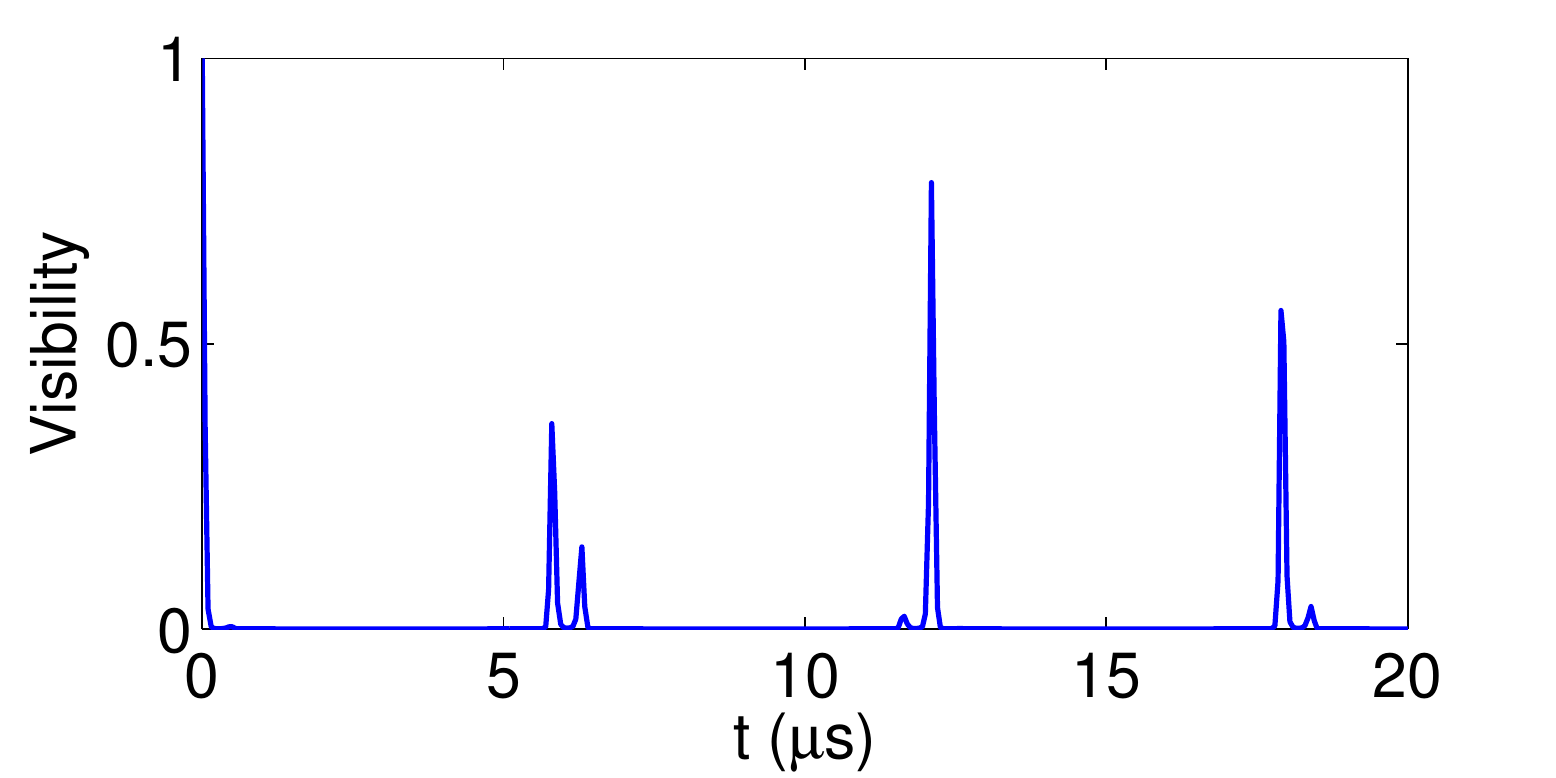}\put(-200,75){(c)}}
  \caption{(Color online) Visibility signal as a function of the time $t$ elapsed between the Ramsey pulses for three ${}^9\mathrm{Be}^+$ ions with an axial trap frequency of $\nu_x=2\pi \times 1 \, \mathrm{MHz}$ and transverse frequencies: (a)~$\nu_y = 2\pi \times 1.565 \, \mathrm{MHz}$, (b)~$\nu_y = 2\pi \times 1.545 \, \mathrm{MHz}$, (c)~$\nu_y = 2\pi \times 1.470 \, \mathrm{MHz}$. The critical value is $\nu_c = 2\pi \times 1.549 \, \mathrm{MHz}$. The frequency of the transverse potential for the central ion when it is excited is $\sqrt{\nu_y^2+\nu_{{\rm dip},y}^2}$, where $\nu_{{\rm dip},y}=2\pi \times 245 \, \mathrm{kHz}$.}
  \label{fig:signals}
\end{figure}

Figure~\ref{fig:signals} displays the visibility of the interferometric signal, given by Eq.~\eqref{eq:overlapmodulus}, as a function of the time elapsed between the two pulses. The visibility is evaluated using the formula derived in Sec.~\ref{sec:harmonic_model}. The three plots correspond to three regimes we consider in this paper. In~\ref{fig:signals}(a) the equilibrium configuration of the crystal is always a linear chain, which experiences a tighter potential when the central ion is excited. Therefore the first quench does not change the equilibrium positions but rather the frequencies of the normal modes. The corresponding visibility exhibits sinusoidal oscillations and is close to unity. Figure~\ref{fig:signals}(b) shows the case when the equilibrium configuration is a zigzag if the central ion is in the ground state, while it is a linear chain when the central ion is excited: The visibility decays quickly to zero, in agreement with the theoretical predictions for the decoherence of a spin coupled to an environment close to criticality~\cite{Zanardi,Paz}, but exhibits revivals with different peak heights. Figure~\ref{fig:signals}(c), finally, corresponds to the situation when the equilibrium configuration of the crystal is always a zigzag, which experiences a shallower potential when the central ion is excited. Here, the quench is also associated with a displacement of the equilibrium positions. Similar to case (b), the signal decays and exhibits revivals. The decay of the visibility in (b) and (c) arises from the entanglement of the spin excitation with the crystal degrees of freedom, and the revivals are a signature of quantum coherence that is stored in the whole system. Similar revivals have been experimentally observed for a single trapped ion~\cite{Monroe2012}. They are here an intrinsic property of the many-body system and, as we will argue in the following, are a scalable feature that appears close to criticality.

The details of the model that determine the properties of the overlap integral, and thus of the visibility, are reported in the next Section. We note that in Eq.~\eqref{eq:overlap}  we assumed that there is no mechanical effect associated with photon absorption and emission. This is the case when the internal transition is excited by means of a radio-frequency field~\cite{Wunderlich}, or by a Raman transition with co-propagating beams~\cite{Leibfried}. The mechanical effects can also included in our formalism, see for instance Ref.~\cite{DeChiara2008}, but will not change substantially the results for the cases illustrated in Figs.~\ref{fig:signals}(b) and (c). If the equilibrium structure is a linear chain independently of the internal state of the ion, as in Fig.~\ref{fig:signals}(a), a momentum kick would induce an oscillation about the equilibrium positions that would modify the signal. We refer the reader to Ref.~\cite{DeChiara2008}, where a similar situation was studied.

\section{Theoretical model}
\label{sec:harmonic_model}
We now give the detailed form of the Hamiltonian $H$ that determines the evolution of the system. In the following we will restrict the motion of the crystal to the $x$-$y$ plane assuming a tight confinement in the $z$ direction, so that the motion along this axis can be considered frozen out. The coordinates ${\bf r}=(x,y)$  give the position in the plane $z=0$. This assumption is made for convenience: The calculations of this paper can be straightforwardly extended to three dimensions.
\subsection{Hamiltonian}
We first consider the internal degrees of freedom. We shall assume that only the central ion can be excited, while all other ions remain always in the ground state. The internal dynamics between the pulses can be restricted to the central ion, with Hamiltonian:
\begin{equation}
H_{\mathrm{el}} = \hbar \omega_{eg} \ketbra{e}{e}, 
\end{equation}
where $\omega_{eg} = \omega_e - \omega_g$ is the transition frequency. The pulses are applied at time $t=0$ and $t$ and correspond to a unitary operation given by the Pauli matrix $\sigma_x$. 

The Hamiltonian for the external degrees of freedom of the ions, $H_{\mathrm{mot}} $, depends on the internal state of the central ion. We denote by $\vek r_i$ the position and by $\vek p_i$ the canonically conjugate momentum of the ion labelled by $i$. The corresponding energy reads
\begin{equation}
 H_{\mathrm{mot}} = H_{\mathrm{kin}} + V_{\mathrm{pot}} + V_{\mathrm{Coul}} \,,
\end{equation}
where $ H_{\mathrm{kin}} = \sum_{i=1}^{N} \vek p_i^2 / (2m) $ is the total kinetic energy, 
\begin{equation}
 V_{\mathrm{Coul}} = \frac{1}{2} \sum_{i=1}^{N} \sum_{ \substack{l=1 \\ l \neq i}}^{N} \frac{q^2}{4\pi\epsilon_0} \frac{1}{| \vek r_i - \vek r_l |} 
\end{equation}
is the Coulomb repulsion, while the energy associated to the external potential takes the form
\begin{equation}
 V_{\mathrm{pot}} = \sum_{i=1}^{N} V_{\mathrm{trap}} (\vek r_i) +  V_{\mathrm{dip}}(\vek r_{i_c}) \ketbra{e}{e} \,.
\end{equation}
Here, the trap potential is
\begin{equation}
V_{\mathrm{trap}}(\vek r_i) = \frac{m}{2} \left( \nu_x^2 x_i^2 + \nu_y^2 \, y_i^2 \right)
\end{equation}
with $\nu_x$, $\nu_y$ the trap frequencies along the axes $x$, $y$ respectively, and the spatial part of the spin-dependent potential reads 
\begin{equation}
V_{\mathrm{dip}}(\vek r_{i_c}) = \frac{m}{2} \nu_{\mathrm{dip}}^2 \, y_{i_c}^2,
\end{equation}
where the subscript $i_c$ labels the central ion. We assume $\nu_{\mathrm{dip}}$ is small compared to $\nu_y$.
The total Hamiltonian which governs the dynamics between the laser pulses takes then the form 
\begin{equation}
  \label{eq:hamiltonian}
  H = H_{\mathrm{el}} + H_{\mathrm{kin}} +  V_{\mathrm{pot}} + V_{\mathrm{Coul}}\,.
\end{equation}
In particular, the Hamiltonian $H_s=\langle s |H_{\mathrm{mot}} |s\rangle$ determines the dynamics of the external degrees of freedom when the central ion is in the internal state $|s=g,e\rangle$.

\subsection{Spin-dependent crystalline structures}
\label{subsec:impurity}
 
We shall consider that the ions vibrate about their classical equilibrium positions, with displacements from equilibrium that are much smaller than the inter-particle distance~\cite{EschnerJOSAB2003}. This situation can be achieved by laser cooling a hot cloud of ions confined in an electromagnetic trap, e.g. a Paul or Penning trap~\cite{DubinONeil1999,Bollinger2012}.The spin-dependent potential can be an optical potential, like a tightly focussed laser beam propagating along and aligned with the chain axis, as discussed in~\cite{Baltrusch2011}. Since this potential depends on the internal state, so does the crystal equilibrium structure. For a fixed number of ions $N$ the relevant parameters controlling the structure of the crystal are the aspect ratio $\alpha = \nu_y^2 / \nu_x^2$ and the state-dependent shift to the aspect ratio $\alpha_{\mathrm{dip}} = \nu_{\mathrm{dip}}^2 / \nu_x^2$, where we consider that the spin-dependent force steepens the potential for the central ion. When all ions are in the ground state and $\alpha$ is larger than a critical value $\alpha_c(N)$, the linear chain is stable, while at $\alpha_c(N)$ it undergoes a continuous transition to a zigzag~\cite{Birkl1992,Dubin1993,Morigi2004}. We shall assume that $\alpha$ is close to this critical value, so that the equilibrium structure depends on the internal state. To study quenches across the phase transition by exciting the central ion, an accurate knowledge of the dynamical properties of the crystalline structures in both configurations is necessary.

\subsection{Spin-dependent normal modes}

In the following we introduce the notation for the normal modes of the state-dependent ion crystal, using $s = g$, $e$ to indicate the internal state of the central ion. We denote by $\vek r_i^{s}$ the equilibrium position of the $i$-th ion in the crystal for each internal state. A Taylor expansion of the potential $ V_{\rm pot} + V_{\mathrm{Coul}}$ is performed to second order for small displacements $\vek q_i^{s}$ around the equilibrium positions, $\vek q_i^{s}=\vek r_i - \vek r_i^{s}  $. For convenience we will use the notation $ q_{i,x} \mapsto q_j$ with $j = 1,\dotsc , N$ and $ q_{i,y} \mapsto q_j$ with $j = N+1,\dotsc , 2N$. The following relation links the displacements between the crystal with the central ion in state $g$ and $e$:
\begin{equation}
\label{eq:ansatzlink}
r_j = r_j^{g} + q_j^{g} =  r_j^{e} + q_j^{e} \,.
\end{equation}
The Hamiltonian of the crystal conditioned to whether the central ion is in state $s=g,e$ takes the form
\begin{equation}
  \label{eq:quadratic_H}
 H^{(s)}_{\mathrm{eff}} \approx \sum_{j=1}^{2N} \frac{p_j^2}{2m} + \sum_{j,k}^{2N}  \frac{m}{2} \, \mathbf{\bar V}^s_{jk} q_j^s q_k^s \,,
\end{equation}
where $\mathbf{\bar V}^s$ is defined as
\begin{equation}
\mathbf{\bar V}^s_{jk} =\frac{\partial^2}{\partial r_j\partial r_k}\left(V_{\rm pot}^s + V_{\mathrm{Coul}}\right)
\Bigl|_{ \vek r_i^s}
\end{equation}
and $V_{\rm pot}^s =\langle s |V_{\rm pot}|s\rangle$.

Hamiltonian~\eqref{eq:quadratic_H} is transformed into a set of uncoupled oscillators by an orthogonal matrix $\mathbf M^s $ such that:
$$\sum_{jk} \mathbf{M}^{s}_{jl} \mathbf{\bar V}^{s}_{jk}\mathbf{M}^{s}_{kn} =  m\left(\omega^{s}_l\right)^2 \delta_{ln}\,,$$
where $\omega_l^{s}$ are the normal modes frequencies and the corresponding coordinates are related to the original displacements by the transformation $Q_l^{ s} = \sum_k \mathbf M_{kl}^{s} q_k^{s}$, with $l=1,\ldots, 2N$. The second-quantized form of the Hamiltonian is found  introducing annihilation (creation) operators $b^{s}_j$ ($b^{s}_j{}^\dagger $), with $b^{s}_j = \sqrt{m \omega^{s}_j /(2\hbar)} \, [Q^{s}_j + {\rm i} P^{s}_j /(m \omega^{s}_j)]$ and $[b_j^{s},b_l^{s\dagger}]=\delta_{jl}$:
\begin{equation}
\label{eq:ext_spin-boson}
 H^{(s)}_{\mathrm{eff}}  = \sum_{s = g,e} \sum_{j=1}^{2N} \ketbra{s}{s} \hbar \omega^{s}_j \left( b^{s}_j{}^\dagger b^{s}_j+\frac{1}{2}\right) \,.
\end{equation}
The eigenstates are the number states $\{ \ket{n_1, \dots, n_{2N}}_{s} \}$ with $b^{s}_j{}^\dagger b^{s}_j\ket{n_1, \dots, n_{2N}}_{s}=n_j\ket{n_1 \dotsb n_{2N}}_{s}$ and $n_j=0,1,2,\ldots$, which form a complete and orthonormal basis for fixed $s$. The eigenstates of $H^{(g)}_{\mathrm{eff}}$ and $H^{(e)}_{\mathrm{eff}}$ are related by a transformation which is specified below and will be needed in order to study the dynamics of the system after the quench. 

\subsection{Mapping between the normal modes of two  different crystalline structures}

In order to evaluate the visibility, which is found from Eq.~\eqref{eq:overlap}, we need to determine the transformation which connects the quantum states of the linear and zigzag structures. In this subsection we show that this is simply found from the transformation which connects the ground states of the linear and of the zigzag configuration. This transformation is derived below, the final result is given in Eq.~\eqref{eq:groundstateoverlap}. 

For this purpose, we first consider the mapping relating the normal modes with displacement $Q_l^{g}$ and $Q_l^{ e}$.  
This is found starting from Eq.~\eqref{eq:ansatzlink} and rewriting it as
\begin{equation}
\label{eq:linkcoords}
q_j^{g} = q_j^{e} + d_j^g \,,
\end{equation}
where $d_j^g = r_j^{e} - r_j^{g}$ is the difference between the equilibrium values for the coordinate $r_j$. Inserting the definition of the normal modes one finds
\begin{subequations}
\label{eq:linkmodes}
 \begin{align}
\label{eq:linkmodes:Q}
Q_j^{g} &= \sum_k \mathbf{T}_{jk} Q_k^{e} + D^g_j  \,,\\
\label{eq:linkmodes:P}
P_j^{g} &= \sum_k \mathbf{T}_{jk} P_k^{e} \,,
\end{align}
\end{subequations}
with $\mathbf{T}_{jl}=\sum_k \mathbf{M}_{kj}^{g} \mathbf{M}_{kl}^{e}$ an orthogonal matrix and $D_j^g = \sum_k \mathbf{M}_{kj}^{g} d_k^g$ the mode displacements. The transformation of the corresponding normal-mode annihilation and creation operators is given by a Bogoliubov transformation, obtained by inserting the definitions of the operators into relations~\eqref{eq:linkmodes}, and which takes the form:
\begin{subequations}
\label{eq:bogoliubov}
 \begin{align}	
\label{eq:bogoliubov:a}
b_j^{g}{}^{\phantom\dagger} &= \sum_k u_{jk} b^{e}_k{}^{\phantom\dagger} - \sum_k v_{jk} b^{e}_k{}^\dagger + \beta^g_j \,,\\
\label{eq:bogoliubov:c}
b_j^{g}{}^\dagger &= \sum_k u_{jk} b^{e}_k{}^\dagger - \sum_k v_{jk} b^{e}_k{}^{\phantom\dagger} + \beta^g_j \,.
\end{align}
\end{subequations}
Here, the real dimensionless coefficients $u_{jk}$, $v_{jk}$ read:
\begin{subequations}
\label{eq:bogoliubov:coeffs}
 \begin{align}
\label{eq:bogoliubov:u}
u_{jk} &= \frac{\mathbf{T}_{jk}}{2}  \left[ \sqrt{\frac{\omega_k^{e}}{\omega_j^{g}}} + \sqrt{\frac{\omega_j^{g}}{\omega_k^{e}}} \right]  \,,\\
\label{eq:bogoliubov:v}
v_{jk} &= \frac{\mathbf{T}_{jk}}{2}  \left[ \sqrt{\frac{\omega_k^{e}}{\omega_j^{g}}} - \sqrt{\frac{\omega_j^{g}}{\omega_k^{e}}}\right] \,,
\end{align}
\end{subequations}
and fulfill the equations~\cite{Fetter}
\begin{subequations}
  \label{eq:bogoliubov:relations}
  \begin{align}
    &\sum_k \left(u_{jk}u_{lk} - v_{jk}v_{lk}\right) = \delta_{jl} \,, \label{eq:bogoliubov:relations:1}\\ 
    &\sum_k \left(u_{jk}v_{lk}-v_{jk}u_{lk}\right) = 0 \quad \forall j,l \label{eq:bogoliubov:relations:2} \,.
  \end{align}
\end{subequations}
Coefficient $\beta^g_j$ describes a displacement of the corresponding normal mode:
\begin{equation}
\label{eq:betadisplacements}
\beta^g_j= \sqrt{\frac{m \omega_j^{g}}{2\hbar}} \, D^g_j\,.
\end{equation}

After having obtained these relations we can now identify the transformation connecting the basis states  $\{ \ket{n_1,n_2,\ldots}_{e} \}$ and  $\{ \ket{n_1,n_2, \ldots}_{g} \}$. Since every state of the bases can be generated from the corresponding ground state by applying repeatedly the corresponding creation operators, it is sufficient to find a mapping between the ground states $\ket{0,0,\dots,0}_{e}\equiv\ket 0_{e}$ and $\ket{0,0,\dots,0}_{g}\equiv\ket 0_{g}$. Such mapping is given by a unitary transformation $ \mathcal U$ such that
\begin{equation}
 \label{eq:unitaryansatz}
 \ket 0_{g} = \mathcal U \ket 0_\mathrm{e} \,.
\end{equation}
Operator $ \mathcal U$ connects two Gaussian states and can thus be written as:
\begin{equation}
\mathcal U = \mathcal D(\gamma_1,\dotsc,\gamma_{2N}) \; \mathcal S (\xi_{11}, \xi_{12}, \dotsc, \xi_{2N\,2N})\,,
\end{equation}
where $\mathcal D$ is a displacement operator and $\gamma_j$ are real scalars, while $\mathcal S$ is a squeezing operator that takes the form
\begin{equation}
\label{eq:Squeez}
\mathcal S = \exp \Bigg( \frac{1}{2}\sum_{jk} \xi_{jk} b^{e}_j{}^\dagger  b^{e}_k{}^\dagger -\xi^*_{jk} b^{e}_j{} b^{e}_k{} \Bigg) \,,
\end{equation}
with squeezing parameters $ \xi_{jk} $ to be determined. With the help of the disentangling theorem, Eq.~\eqref{eq:Squeez} can be recast into the convenient form 
\begin{equation}
\label{eq:Squeez:1}
\mathcal S = Z e^\mathrm{A} e^\mathrm{B} e^{-\mathrm{A}^\dagger} \,,
\end{equation}
where $ Z $ is a scalar while
\begin{subequations}
\label{eq:disent}
 \begin{align}
\label{eq:disent:A}
\mathrm A &= \frac{1}{2}\sum_{jk} A_{jk} \, b_j^{e}{}^\dagger b_k^{e}{}^\dagger \,, \\
\label{eq:disent:B}
\mathrm B &= -\sum_{jk} B_{jk} \, b_j^{e}{}^\dagger b_k^{e}{} \,,
\end{align}
\end{subequations}
are operators, with $A_{jk}$ a symmetric matrix. Details of the derivation are provided in Appendix~\ref{sec:disentangling}. Application of operator~\eqref{eq:Squeez:1} to the state $|0\rangle_e$ gives
\begin{equation}
  \label{eq:Fetter:transformation}
 \ket 0_{g} = \mathcal W \ket 0_{e} \,,
\end{equation}
with the non-unitary operator $\mathcal W$ defined as:
\begin{equation}
 \mathcal W = Z \, \mathcal D(\gamma_1,\dotsc,\gamma_{2N}) \, e^\mathrm A \,.
\end{equation}
We note that this operator was first introduced in~\cite{FetterAoP72} for evaluating the thermodynamics of interacting condensates. 

We now determine the coefficients $A_{jk}$, the displacements $\gamma_j$ and the normalization constant $Z$.
For this purpose we make use of relation $ b_j^{g} \ket 0_{g}=0$ which must hold for any mode~$j$ of $H_{\rm eff}^{(g)}$. Using Eqs.~\eqref{eq:bogoliubov:a} and~\eqref{eq:Fetter:transformation}, one obtains:
\begin{equation}
\label{eq:ansatz:Fetter}
0 = b_j^{g} \ket 0_{g} 
  =  \left[ \sum_k \left(  u_{jk} b_k^{e}- v_{jk} b_k^{e}{}^\dagger \right) + \beta^g_j \right] \mathcal{W} \ket 0_{e} \,,
\end{equation}
which can be recast in the form 
\begin{multline}
\label{Eq:W}
 \mathcal W \left\lbrace \biggl[ \sum_{k}  u_{jk} \Bigl( \sum_l A_{kl} b_l^{e}{}^\dagger \Bigr) - \sum_k v_{jk} b_k^{e}{}^\dagger \biggr] \right. \\
  +\left.  \biggl[ \sum_k \left( u_{jk} -v_{jk} \right) \gamma_k + \beta^g_j \biggr] \right\rbrace \ket 0_{e} =0 
\end{multline}
The latter equation has been derived from (\ref{eq:ansatz:Fetter}) multiplying by $\mathds{1} = \mathcal W \mathcal W^{-1}$ on the left side and making use of the relations 
\begin{subequations}
     \begin{align}
      \mathcal W^{-1} b_k^{e}\phantom{{}^\dagger} \mathcal W &= b_k^{e}\phantom{{}^\dagger} + \sum_l A_{kl} b_l^{e}{}^\dagger + \gamma_k \,,\\
      \mathcal W^{-1} b_k^{e}{}^\dagger \mathcal W &= b_k^{e}{}^\dagger + \gamma_k \,.
    \end{align}
\end{subequations}
Equation~\eqref{Eq:W} is equivalent to:
\begin{eqnarray}
&&
    \sum_k u_{jk} A_{kl} -v_{jl} = 0 \, \label{eq:Aeq}\\
&&   \label{eq:gamma}
\sum_k (u_{jk}-v_{jk})\gamma_k + \beta^g_j = 0 \label{eq:betaeq}\,,
\end{eqnarray}
which must hold for all $j,l=1,\ldots, 2N$. From Eq. (\ref{eq:Aeq}) one finds the coefficients
\begin{equation}
\label{eq:mapping:A}
A_{jk}= \sum_l (u^{-1})_{jl} \, v_{lk}\,, 
\end{equation}
where one sees that $A$ is real, with symmetry following from (\ref{eq:bogoliubov:relations:2}), while from Eq.~\eqref{eq:betaeq} obtains
\begin{equation}
  \label{eq:mapping:beta^e}
  \gamma_j= - \sum_k (u_{kj} + v_{kj}) \beta^g_k =: \beta^e_j\,,
\end{equation}
where $\beta^e_j$ has been defined. Finally, the constant $Z$ is found from the condition that the norm of the ground state $ | 0 \rangle_{{g}}$ must be unity, ${}_{{g}}\langle 0 | 0 \rangle_{{g}}=1$, giving \begin{align}
1=\phantom{}_{{e}}\langle 0 |\,\mathcal W^\dagger \mathcal W \ket{0}_{{e}}= Z^2 \phantom{}_{e}\langle 0 |\, \left( \sum_{n=0}^\infty \sum_{m=0}^\infty \frac{{\mathrm{A}^{\!\dagger}}^n {\mathrm A^{\vphantom \dagger}}^m}{n! \, m!} \right)  \ket{0}_{e} \,, \label{eq:normalization}
\end{align}
which leads to:
\begin{equation}
  \label{eq:mapping:Z}
  Z = \det \left[ \left( 1-\mathrm A^2 \right)^{1/4} \right]
\end{equation}
(details are given in Appendix~\ref{sec:linkedcluster}). 

Using this result, we find:
\begin{equation}
\label{eq:trafo}
 \ket{ 0}_g = Z \, \mathcal D_e(\beta^e) e^{\mathrm{A}} \ket{ 0}_e\,,
\end{equation}
with $A_{jk}$, $\beta^e_j$, and $Z$ given in Eqs.~\eqref{eq:mapping:A},~\eqref{eq:mapping:beta^e} and~\eqref{eq:mapping:Z}, respectively. Using this formalism we can now evaluate the visibility (which we defer for Sec.~\ref{sec:visibility}) as well as the overlap between the two ground states:
\begin{align}
 \label{eq:groundstateoverlap}
  \mathcal G_0 &= \braketIX{e}{0}{0}{g} = Z \matrixElementSub{0}{e}{\mathcal D(\beta^e)e^{\mathrm{A}}}{0}{e} \nonumber \\
	     &= Z \exp \Bigg\{ \frac{1}{2}\sum_{jk} A_{jk}\beta^e_j\beta^e_k \Bigg\} \exp \Bigg\{\!-\frac{1}{2} \sum_j |\beta^e_j|^2 \Bigg\} \,.
\end{align}
Before we conclude this section, we also give the form of the squeezing parameters $\xi_{jk}$ in operator $\mathcal S$ (\ref{eq:Squeez:1}):
\begin{equation}
\label{xi}
 \xi_{jk} = \sum_{l} \Lambda_{jl} \atanh (a_l) \Lambda_{kl} \,,
\end{equation}
where $\Lambda_{jl}$ is the orthogonal transformation diagonalizing $A$ and $a_l$ are the corresponding eigenvalues. The parameters $\xi_{jk}$ are real, since $\mathrm{A}_{jk}$ is real and symmetric. The derivation of Eq.~\eqref{xi} can be found in Appendix~\ref{sec:disentangling}.

\subsection{Evaluation of the visibility}
  \label{sec:visibility}

We now derive an analytical expression for the visibility. Our starting point is the overlap as a function of the time $t$ between the pulses, as given in Eq.~\eqref{eq:overlap}. Using $\ket{\phi}_s=\exp\{H_{\mathrm{eff}}^{(s)}t/{\mathrm{i}}\hbar\}\ket{0}_g =: U_s (t) \ket{0}_g$, we rewrite it as 
\begin{equation}
  \mathcal O(t) = \matrixElementSub{0}{g}{U^\dagger_g(t) U_e(t)} {0}{g} 
  =\matrixElementSub{ 0}{g}{U_e(t)}{0}{g} \,.  \label{eq:overlap1}
\end{equation}
We note that this expression is given up to a time-dependent phase, which depends on the difference between the (classical) ground-state energies of the two equilibrium configurations~\cite{Fishman2008}. Since this factor is irrelevant for the visibility, it will be omitted from now on. Using Eq.~\eqref{eq:trafo}, expression~\eqref{eq:overlap1} can be cast in the form:
\begin{eqnarray}
  \mathcal O(t) &=& Z^2 {}_e\!\matrixElement{0}{e^{\mathrm{A}^\dagger}\mathcal D_e^\dagger (\beta^e) U_e(t) \mathcal D_e (\beta^e)e^{\mathrm{A}}}{0}_e \nonumber\\
&=& Z^2 {}_e\!\matrixElement{\beta^e}{ e^{{\tilde{\mathrm{A}}}^\dagger(\beta^e)}  U_{e}(t)   e^{\tilde{\mathrm{A}}(\beta^e)} }{\beta^e}_{e} \,,\label{eq:O:2}
\end{eqnarray}
where $\ket{\beta^e}_e= {\mathcal D_e(\beta^e) }\ket{0}_e$ and from the first to the second line we employed the relation 
\begin{equation}
\mathcal D_e (\beta^e) e^{\mathrm{A}} = e^{\tilde{\mathrm{A}}(\beta^e)}  \mathcal D_e (\beta^e) \,,
\end{equation}
with
\begin{equation}
  \tilde{\mathrm{A}}(\beta) = \frac{1}{2} \sum_{jk} A_{jk} (b_j^e{}^\dagger - \beta_j^*)(b_k^e{}^\dagger - \beta_k^*) \,.
\end{equation}
%
%
%
Using the overcompleteness of the multimode coherent states, the identity operator reads $\mathds{1} = \frac{1}{\pi^{2N}} \int \mathrm{d}^2{\alpha}_1 \ldots \mathrm{d}^2{\alpha}_{2N} \ketbraIX{\alpha}{e}{\alpha}$, and Eq.~\eqref{eq:O:2} takes the form
\begin{align}
  \mathcal O(t) = \frac{Z^2}{\pi^{2N}} & \int \mathrm{d}^2{\alpha}_1 \ldots \mathrm{d}^2{\alpha}_{2N} \nonumber \\
&  {}_e\!\matrixElement{ \beta^e}{ e^{{\tilde{\mathrm{A}}^\dagger}( \beta^e)} U_e(t) }{{\alpha}}_e \! \matrixElement{{\alpha}}{ e^{\tilde{\mathrm{A}}(\beta^e)}  }{\beta^e}_e \nonumber\\
=\frac{Z^2 }{\pi^{2N}} & \int \mathrm{d}^2{\alpha}_1 \ldots \mathrm{d}^2{\alpha}_{2N}  \nonumber\\
&   e^{[f(\alpha(t)-{\beta^e})]^*} e^{f( \alpha- \beta^e)} 
 \mathcal C_{\beta^e,\alpha(t)}\mathcal C_{\alpha,\beta^e} \,.
\label{eq:O:3}
\end{align}
Here, we have defined 
\begin{equation}
  \label{eq:coherentstateoverlap}
 \mathcal C_{\alpha,\beta} = {}_e\!\braket{ \alpha}{\beta}_e =  e^{  - \sum_j \big( \frac{\modulus{\alpha_j}^2}{2} + \frac{\modulus{\beta_j}^2}{2} - \alpha_j^* \beta \big)  }\,,
\end{equation}
and used 
\begin{equation}
  {}_e\!\matrixElement{ \beta^e}{ e^{{\tilde{\mathrm{A}}^\dagger({\beta^e})}}  }{{\alpha}(t)}_e 
  = e^{[f(\alpha(t)-{\beta^e})]^*} {}_e\!\braket{{\beta^e}}{\alpha(t)}_e \,,
\end{equation}
where $ \ket{\alpha_j(t)} = \ket{\alpha_j \exp\{-{\mathrm{i}}\omega_j^et\}} $ is the time-evolved coherent state and $ f({\alpha}) = \frac{1}{2} \sum_{jk} A_{jk} \alpha_j^* \alpha_k^* $ a $ 2N $-dimensional complex-valued function. The evaluation of Eq.~\eqref{eq:O:3} is just a matter of algebra and is shown in Appendix~\ref{sec:overlapintegral}.
The result reads 
\begin{equation}
 \label{eq:overlap:groundstate:result2}
  \mathcal O(t) = \frac{e^{\frac{1}{4} \vek w^T \Omega^{-1} \vek w}}{\sqrt{\det \Omega}}  \left| \mathcal G_0 \right|^2 \,,
\end{equation}
where the complex symmetric $4N\times4N$ matrix $\Omega$ and the $4N$-dimensional vector  $\vek w$ are defined as
\begin{align}
  \Omega &= \begin{pmatrix} 1 - \Lambda^+ & \;\, -{\mathrm{i}}\Lambda^- \\ -{\mathrm{i}}\Lambda^- & 1 + \Lambda^+ \end{pmatrix} 
  \,,      &
  \vek w &= \begin{pmatrix}
                 \phantom{-{\mathrm{i}}}S^+ \\ -{\mathrm{i}}  S^- 
            \end{pmatrix} .
\end{align}
Here, 
\begin{subequations}
  \label{eq:overlap:groundstate:quantities}
  \begin{align}
    \Lambda_{jk}^\pm &=  \frac{1}{2} A_{jk} \big[ e^{-{\mathrm{i}}(\omega_j^e+\omega_k^e)t} \pm 1 \big]  \,, \\
    S^\pm_j &= S_{j}(\beta^e{}^*) \pm S_{j}(\beta^e) e^{-{\mathrm{i}}\omega_j^e t} \,,
  \end{align}
\end{subequations}
and
\begin{equation}
  S_j(\beta) = \sum_{k}  A_{jk} \beta_k -   \beta_j^* \,.
\end{equation}
Equation~\eqref{eq:overlap:groundstate:result2} determines the visibility, that is plotted in Sec.~\ref{sec:results} for several parameter regimes in which this formula is valid. 

The short-time behaviour of the visibility is found by performing a Taylor expansion of (\ref{eq:overlap1}), and reads:
\begin{equation}
\label{eq:V:tshort}
 \mathcal V(t)\approx 1 + \eta t^2 / 2 \,.
\end{equation}
Quantity $\eta < 0$, denoted in Sec.~\ref{sec:results} as the curvature, is determined by the variance of $H_{\rm eff}^{(e)}$ in the initial state:
\begin{equation}
 \eta = -\frac{1}{\hbar^2} \left[ {}_g\!\bra{0} \big( H_{\rm eff}^{(e)} \big)^2 \ket{0}_g - \big( {}_g\!\bra{0} H_{\rm eff}^{(e)} \ket{0}_g \big)^2 \right].
\end{equation}
The functional dependence of $\eta$ on the parameters in Eq.~\eqref{eq:overlap:groundstate:result2} is derived and given in Appendix~\ref{sec:overlapintegral}.

\section{Quantum quenches at the linear-zigzag transition}
  \label{sec:results}

\begin{figure}[tbhp]
  \centering
  \includegraphics[width=0.4\textwidth]{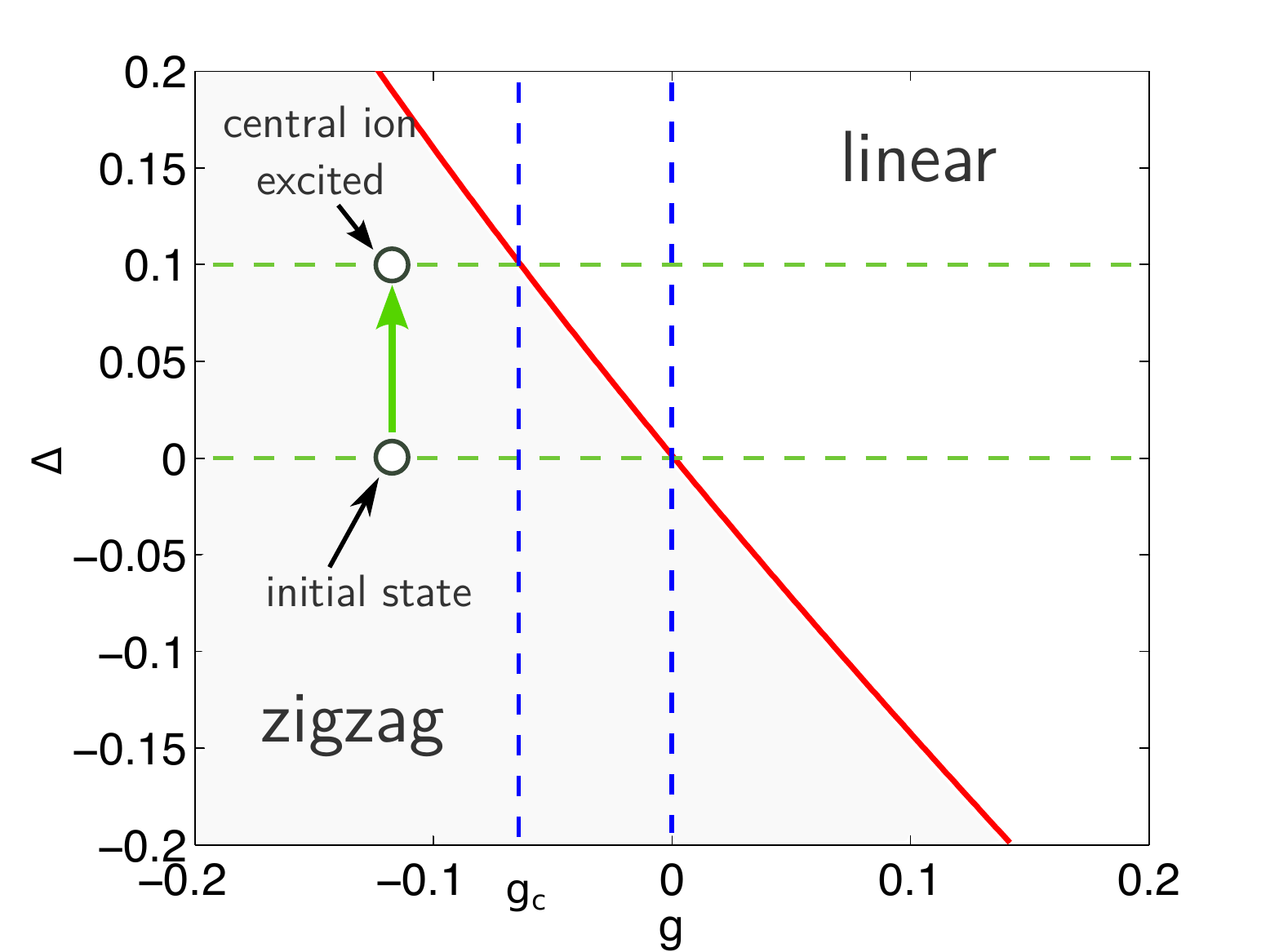}
  \caption{(Color online) Phase diagram for three ions as a function of the dimensionless parameters $g$ and $\Delta$. The solid red line separates the parameter regions where the ions form a zigzag (bottom left) or a linear chain (top right). The crystalline structure corresponding to state $|g\rangle$ is at $\Delta=0$.  Three regimes are considered, depending on where the state before and after the pulse are located in the diagram: when they are both in the left region, $g<g_c$, the chain is in a zigzag structure; when they are both in the right region, $g>0$, the chain is linear, while when they are across the line, $g_c<g<0$, the initial ground state is a zigzag and the final is a linear chain.}
  \label{fig:scheme}
\end{figure}

We shall now examine the visibility of the interferometric signal when the chain is close to the linear-zigzag instability. We assume that a quench is performed by exciting the central ion in presence of a spin-dependent force. 
Due to the long-range interaction, the force on the central ion can induce a change of the equilibrium configuration of the entire crystal. In particular, if $\nu_y^2<\nu_c^2<\nu_y^2+\nu_{\mathrm{dip},y}^2$, the two equilibrium configurations corresponding to the different internal states can be a zigzag and a linear chain, respectively, provided that the correlation length is larger than the size of the system~\cite{Fishman2008,delCampo2010}. The equilibrium configurations corresponding to the central ion excitation are represented in the diagram of Fig.~\ref{fig:scheme}. Here, the horizontal axis gives the dimensionless parameter $g$, which is defined as
\begin{equation*}
  g = \frac{\nu_y^2-\nu_{c}^2}{\nu_{c}^2}\,.
\end{equation*}
This parameter determines whether the equilibrium configuration corresponding to the central ion in state $|g\rangle$ is a linear ($g>0$) or a zigzag chain ($g<0$). The vertical axis gives the dimensionless parameter $\Delta$, defined as
\begin{equation*}
  \Delta =\frac{\nu_{\mathrm{dip}}^2}{\nu_{c}^2}\,,
\end{equation*}
and related with the change in the potential on the central ion when it is in state $|e\rangle$. The equilibrium configuration when the ion is in state $|e\rangle$ is shown in the diagram as a function of $g$ and $\Delta$. We restrict to the case $\Delta>0$, consistently with the choice that the spin-dependent force is restoring. The ion in the excited state feels thus a change of the transverse potential corresponding to a vertical shift in the diagram as sketched by the green arrow. 

Three situations will be discussed in the regime close to the linear-zigzag instability, indicated by the solid line of Fig.~\ref{fig:scheme}. The first one corresponds to the case in which the crystal is initially forming a linear chain ($g>0$). The spin excitation then does not change the equilibrium configuration, nevertheless it modifies the frequencies of the normal modes. An example of the visibility one measures in this case is displayed in Fig.~\ref{fig:signals}(a). When $g<0$, the equilibrium configuration of the initial state is a zigzag. Whether the equilibrium configuration of the excited state is a zigzag or a linear, depends here on whether the shift $\Delta$ is below or above the instability line. For a fixed $\Delta$, this defines a critical value $g_c(\Delta)$, such that at $g=g_c(\Delta)<0$ the crystal equilibrium configuration for the excited state is exactly on the instability line (note that $g_c$ depends on the number of ions $N$). If $g<g_c(\Delta)$, hence, the crystal equilibrium configuration in the excited state is also a zigzag (with however different transverse displacement as the initial one). An example for the visibility found in this case is shown in~\ref{fig:signals}(c). If $0>g>g_c(\Delta)$, instead, the crystalline structures of ground and excited states are a zigzag and a linear chain, respectively. The quench hence drives the chain across the critical point, and a typical visibility signal is shown in Fig.~\ref{fig:signals}(b).
     
\begin{figure}[tbp]
  \centering
   \subfloat{\phantom{(a)}\label{fig:curv:surf}\includegraphics[width=0.4\textwidth]{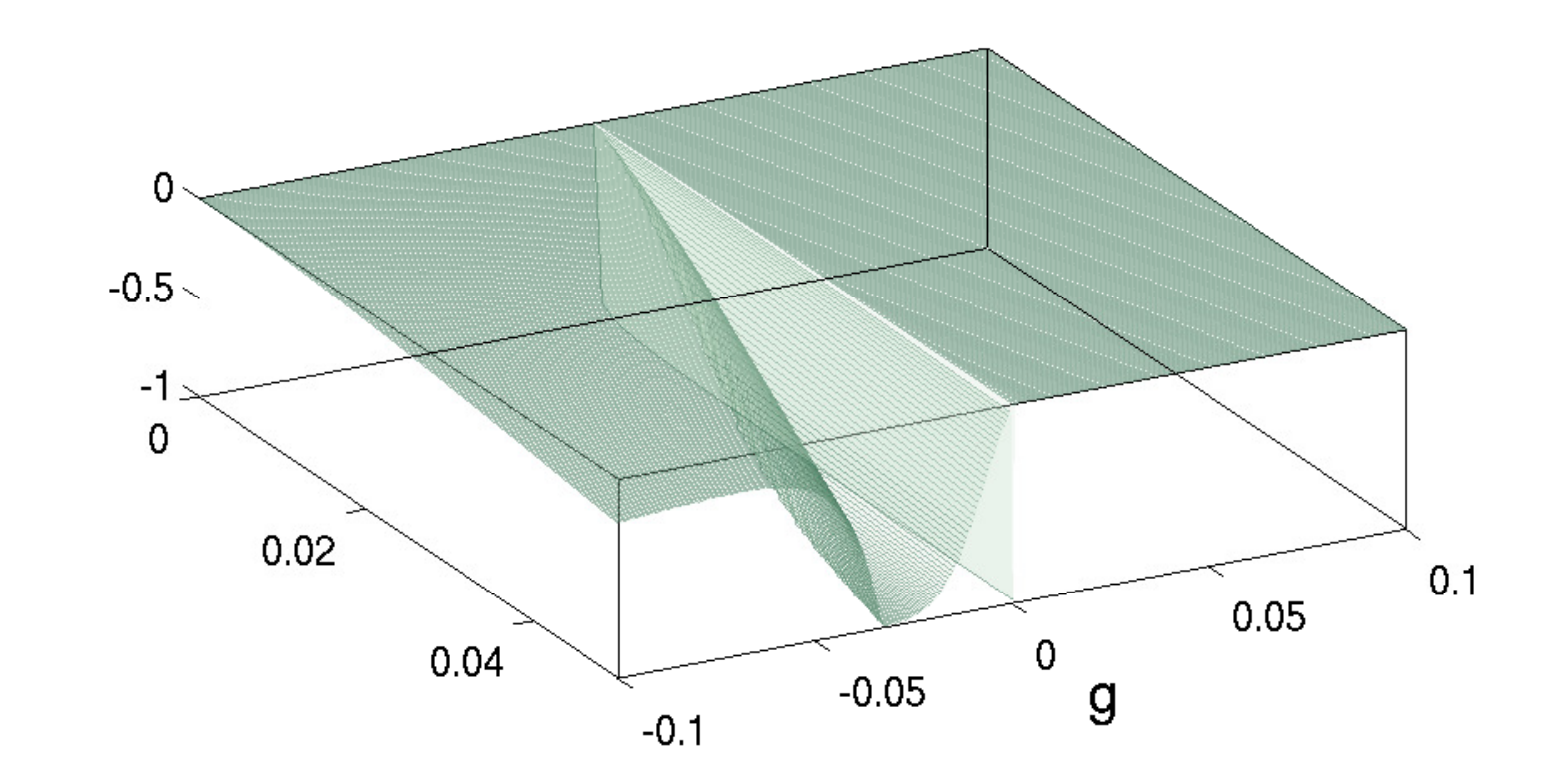}\put(-210,90){(a)}
   \put(-180,15){ $\Delta$}\put(-200,65){$\eta$}}  \\
   \vspace{-.35cm}
   \subfloat{\phantom{(a)}\label{fig:curv:contour}\includegraphics[width=0.4\textwidth]{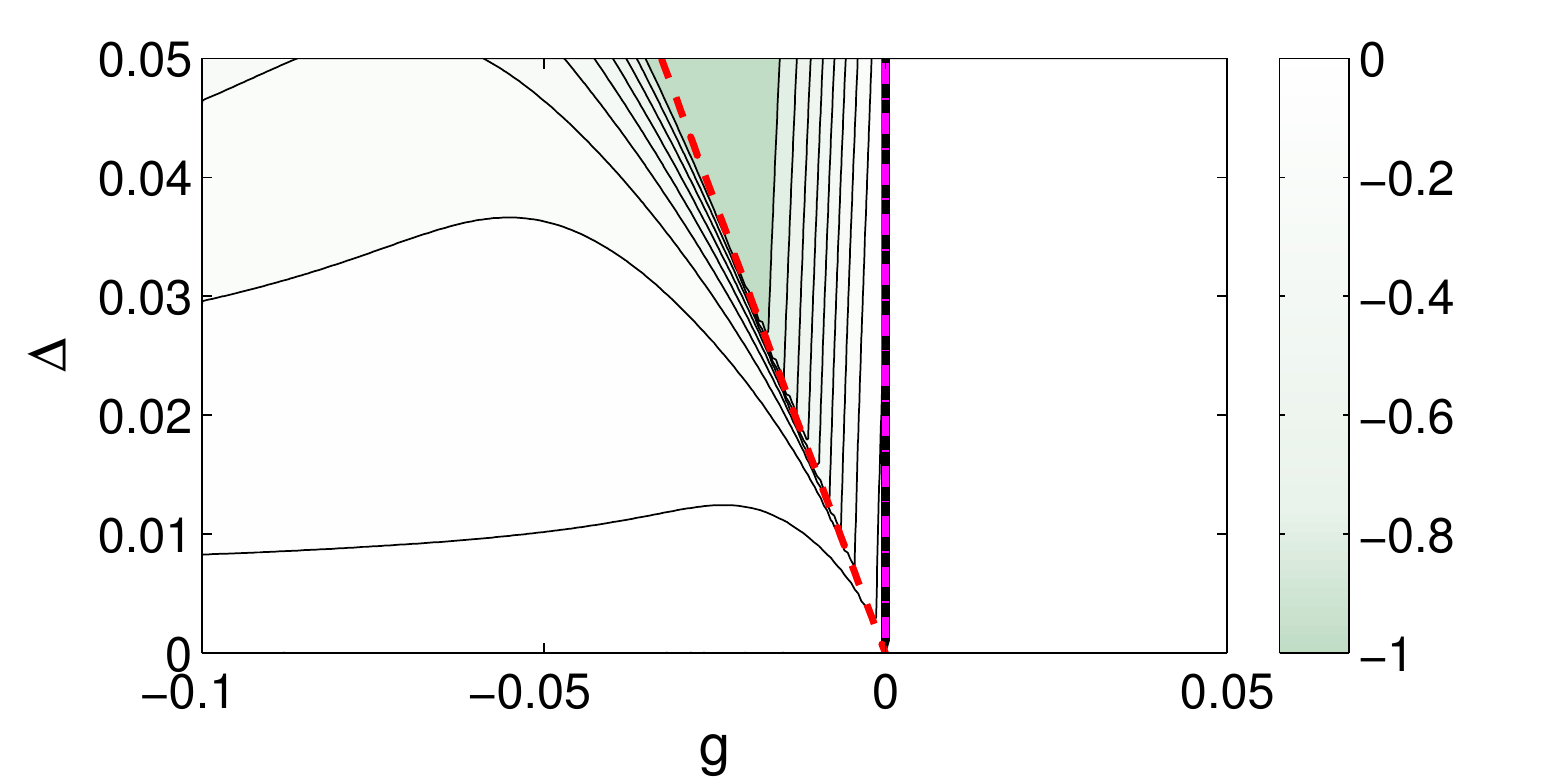}\put(-210,90){(b)}}
\caption{(Color online) Curvature $\eta$ of the visibility signal at short time, Eq.~\eqref{eq:V:tshort},  as a function of $g$ and $\Delta$. Subplot~(b) shows the corresponding contour plot. The curve has been evaluated for three ${}^9\mathrm{Be}^+$ ions with axial trap frequency $\nu_x=2\pi \times 1\mathrm{MHz}$.}
\label{fig:curv}
\end{figure}

The behaviour of the visibility for short times is characterized by a decay with quadratic dependence on the elapsed time, according to Eq.~\eqref{eq:V:tshort}. Figure~\ref{fig:curv}(a) displays the parameter $\eta$ as a function of $g$ and $\Delta$. Decay is faster in the region where the quench is performed across the phase transition, where the overlap between initial and final states is small. The plot is reminiscent of the features of the stability diagram in Fig.~\ref{fig:scheme}, as is visible by inspecting the contour plot in Fig.~\ref{fig:curv}(b).

\begin{figure}[tbp]
  \centering
  \includegraphics[width=0.475\textwidth]{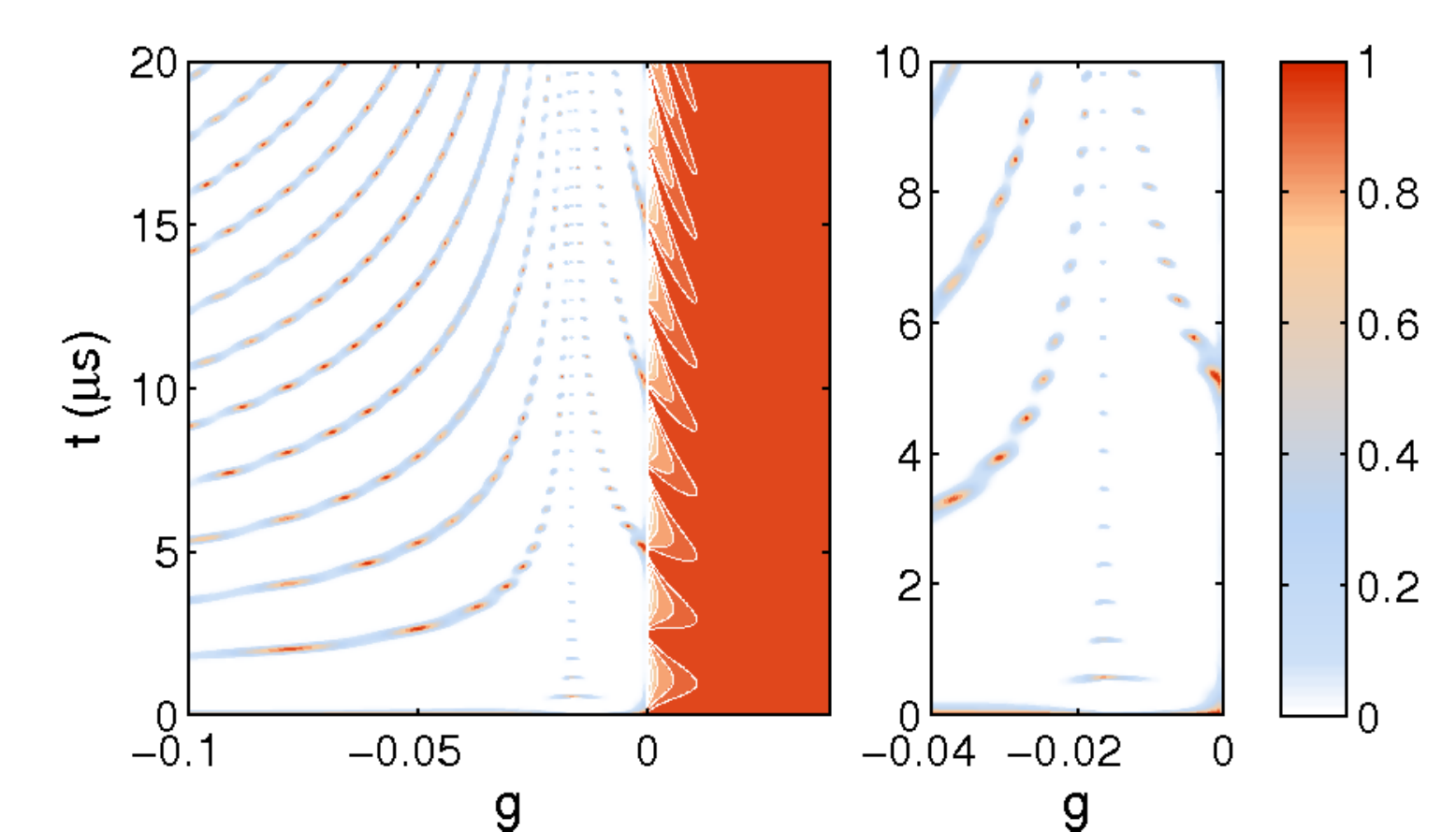}
  \caption{(Color online) Density plot of the visibility signal as a function of $g$ and $t$ (in $\mu$s) for  $\Delta=0.025$. The other parameters are the same as in Fig.~\ref{fig:curv}. The inset enlarges the region about $g_c$  for short elapsed times.} 
  \label{fig:density}
\end{figure}

We now analyse the behaviour for long times. Figure~\ref{fig:density} displays the density plot of the visibility as a function of the rescaled aspect ratio $g$ and of the time $t$ between the pulses. Three distinct behaviours are observed corresponding to the three regimes we identified. For $g<g_c$ the appearance of the revivals is periodic. The corresponding period diverges as $g$ approaches $g_c$. Each curve indicating a maximum of the visibility, moreover, exhibits an additional modulation, showing that the height of the revival peaks varies with $g$. The inset shows the behaviour at $g=g_c$: Here, several peaks of the visibility appear at short elapsed times, with rapidly vanishing height. In the interval $g_c<g<0$ the periodic structure of the revivals is also observed, with decreasing period as $g$ approaches 0. Finally, for $g>0$ the visibility is close to unity and exhibits some modulation for small positive $g$, with an amplitude that vanishes as $g$ increases.

\begin{figure}[tbp]
  \centering
  \includegraphics[width=0.4\textwidth]{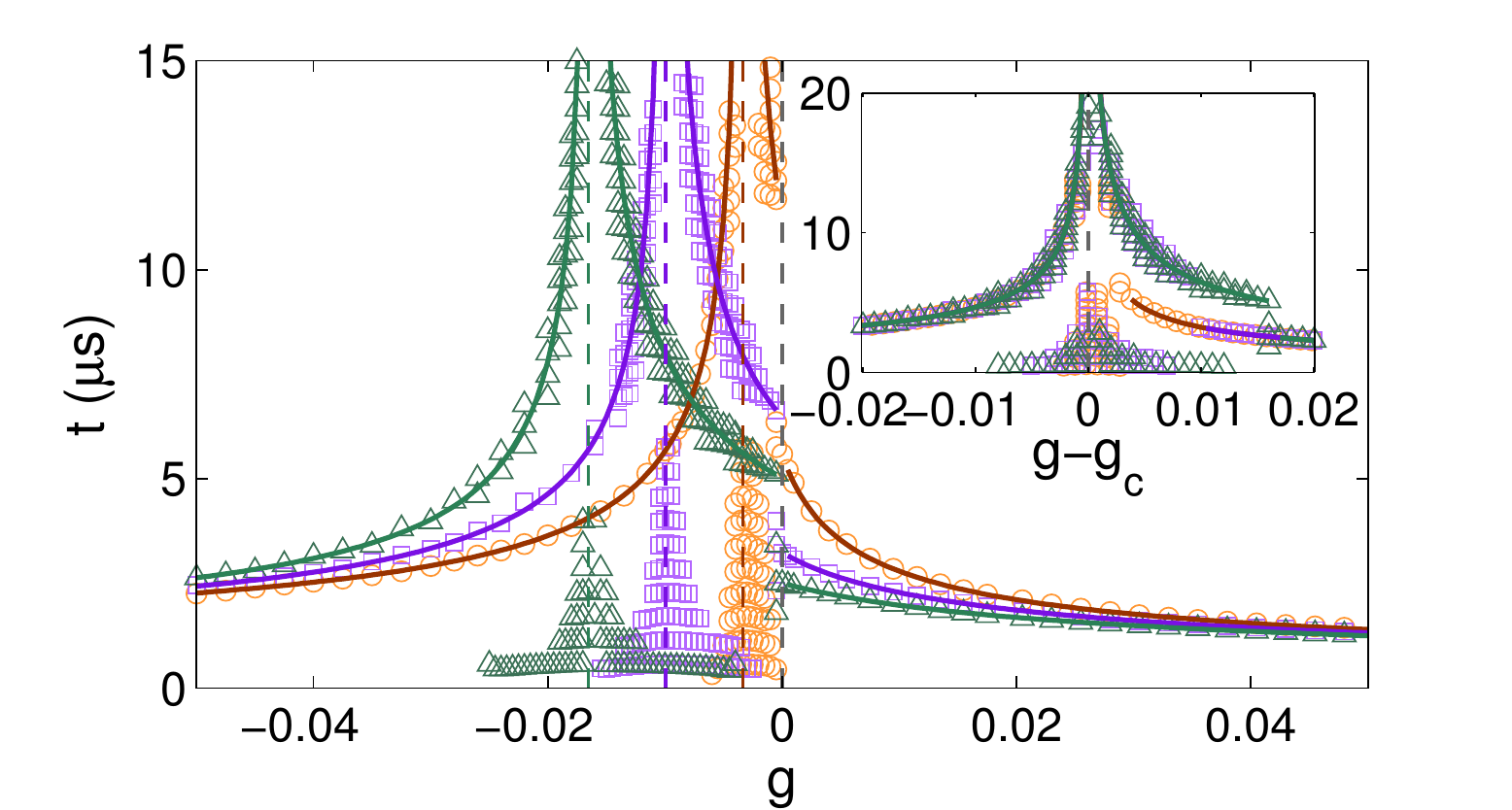}\put(-210,95){(a)} \\
  \includegraphics[width=0.4\textwidth]{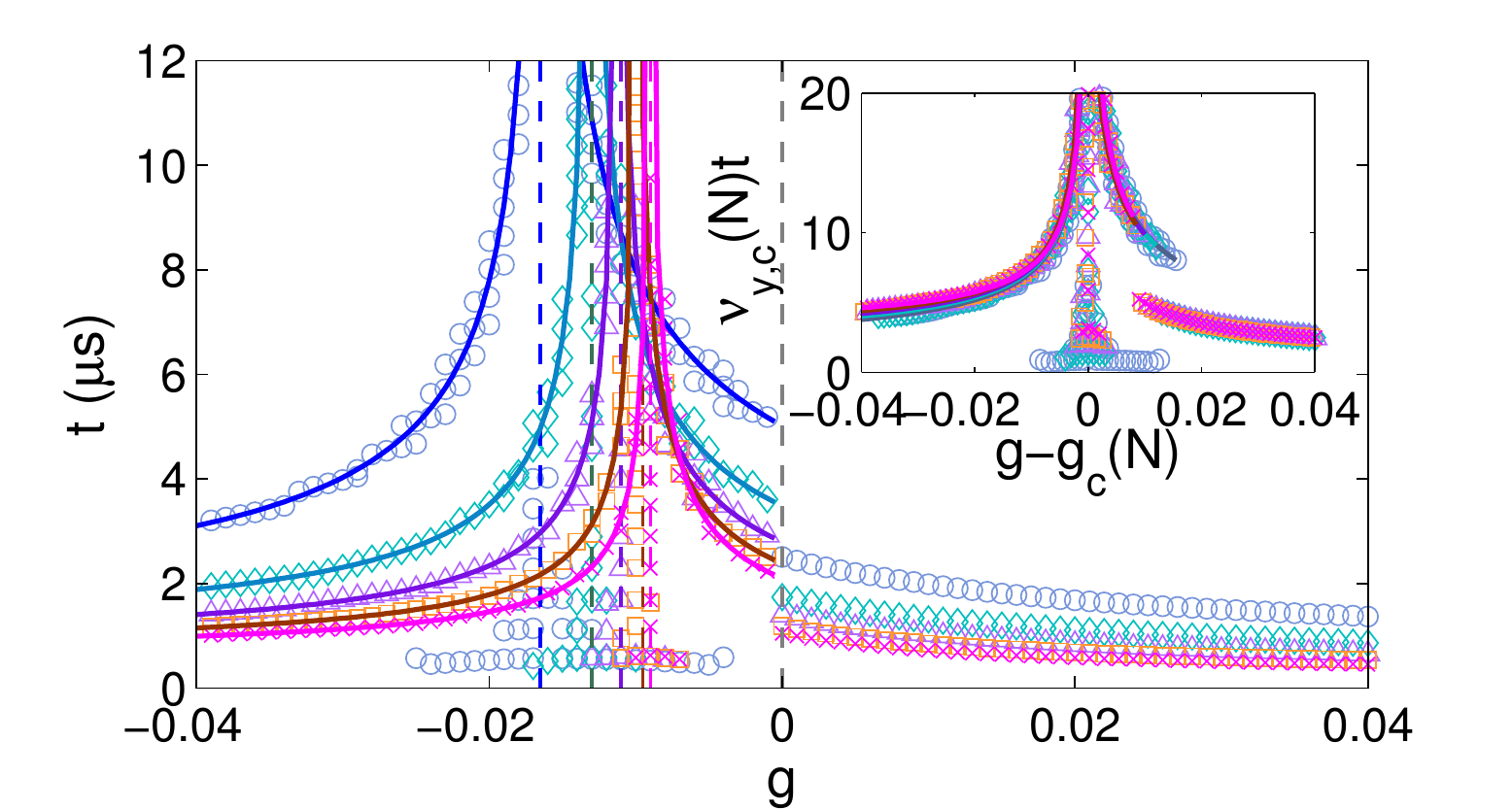}\put(-210,95){(b)}
  \caption{ \label{fig:peaks_3ions_posdelta} (Color online) (a) Times (in $\mu$s) corresponding to the first peak of the visibility as a function of $g$ for different values of $\Delta=0.025$~(symbol:~\textcolor{pine}{$\pmb{\triangle}$}), 0.015 (\textcolor{lavendel}{$\pmb{\square}$}), 0.005 (\textcolor{gold}{$\pmb{\ocircle}$}). The dashed vertical lines denote the different values of $g_c$.  Inset: the same curves are plotted as a function of $g-g_c(\Delta)$. The other parameters are the same as in Fig.~\ref{fig:curv}.
 \label{fig:different_N}  (b) Same as (a) but for a number of ions $N$ = 3 (\textcolor{azure}{$\pmb{\ocircle}$}), 5 (\textcolor{turquoise}{$\pmb{\lozenge}$}), 7 (\textcolor{lavendel}{$\pmb{\triangle}$}), 9 (\textcolor{gold}{$\pmb{\square}$}), 11~(\textcolor{magenta}{$\pmb{\times}$}) with  $\Delta=0.025$. In the inset the curves are plotted as a function of $g-g_c(N)$ with the time in units of $1/\nu_c(N)$.}
\end{figure}

Let us now make some considerations. In the first place, the value of $g_c$ depends on $\Delta$. Nevertheless, the behaviour found in Fig.~\ref{fig:density} is encountered for different values of $\Delta$, as is visible in Fig.~\ref{fig:peaks_3ions_posdelta}, where we show the revival times at which the visibility is different from zero. 
This behaviour is also to large extent independent of the number of ions composing the crystal, as is indicated by Fig.~\ref{fig:different_N}. Here, one observes that all curves giving the first revival time for different numbers of ions exhibit a similar functional dependence as $g$ approaches $g_c$ (which depends also on the number of ions $N$). This behaviour becomes evident by appropriately rescaling the curves as shown in the inset. The width of the peak at $g_c$ in turn decreases as the number of ions is increased. 

\begin{figure*}[tbp]
  \centering
  \subfloat[]{\label{fig:spectrum:lin}\includegraphics[width=0.32\textwidth]{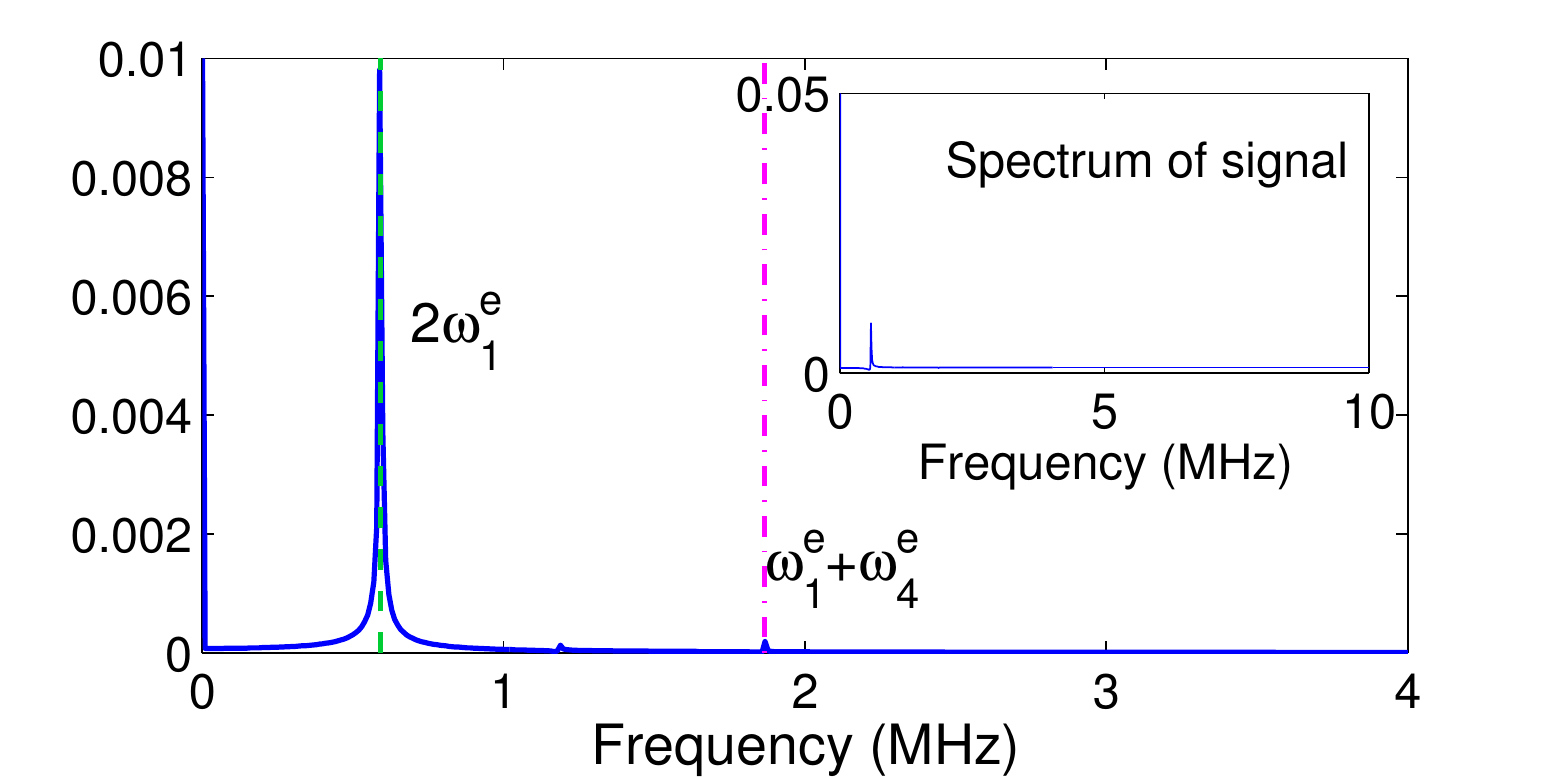}}
\;
\subfloat[]{\label{fig:spectrum:zz}\includegraphics[width=0.32\textwidth]{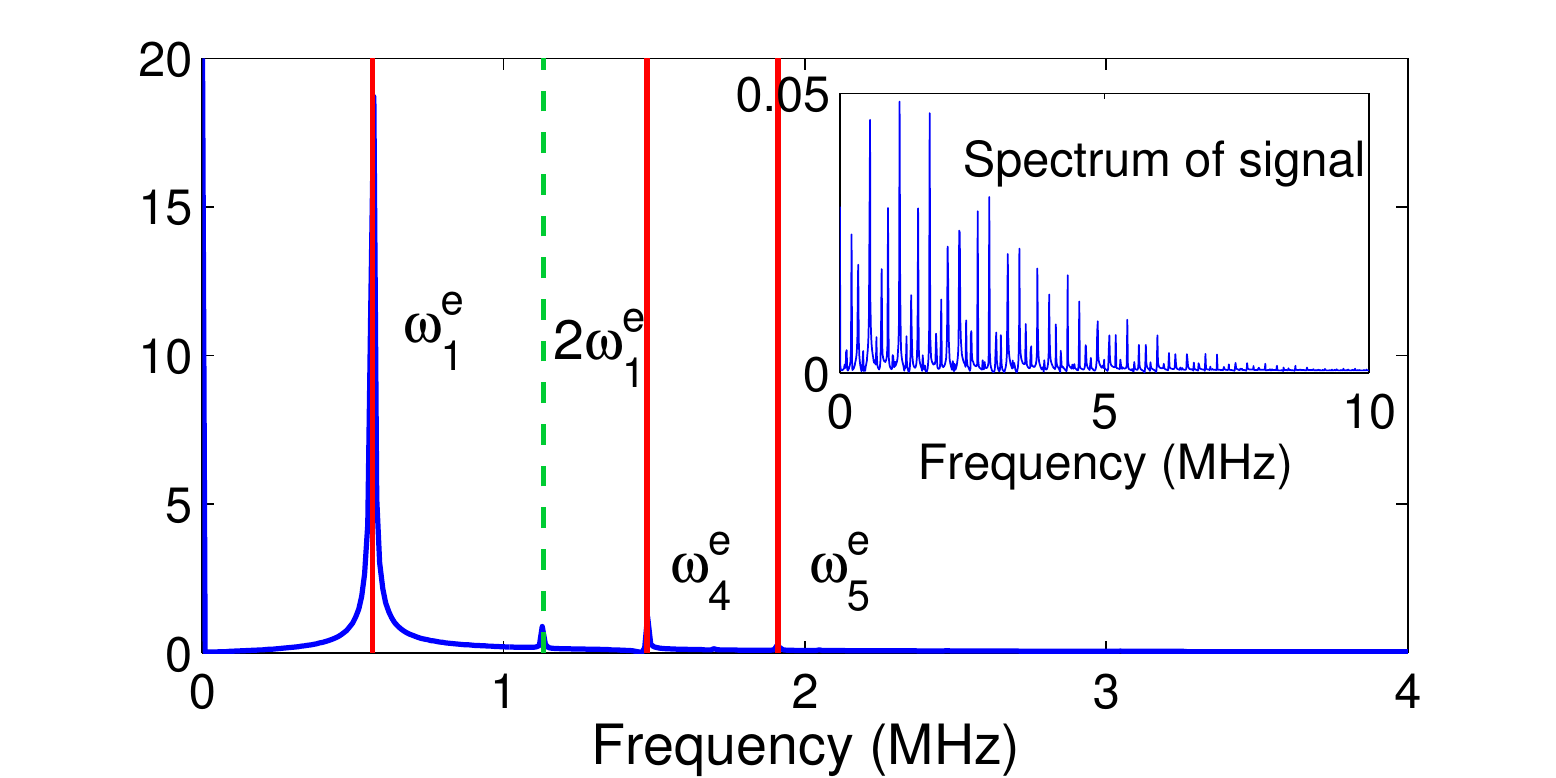}}
\;
  \subfloat[]{\label{fig:spectrum:trans}\includegraphics[width=0.32\textwidth]{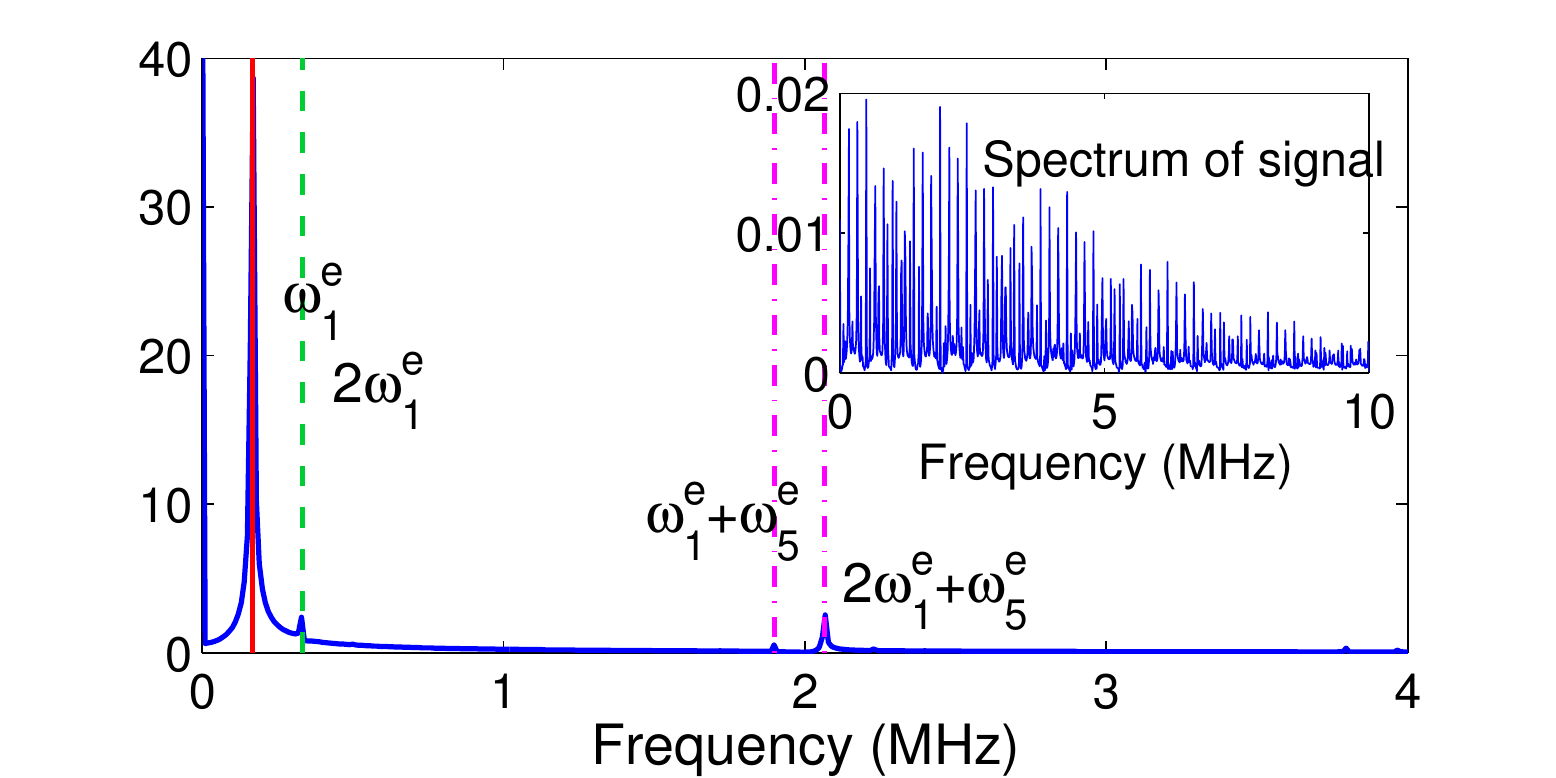}}
\caption{(Color online) The spectrum of the logarithm of the visibility, $\mathcal F_L(\omega_n)$ in Eq.~\eqref{F:log},  corresponding to the
plots in Fig.~\ref{fig:signals}. The vertical straight red lines
indicates the normal-mode frequencies of the crystal when the central
ion is excited, the dashed green lines give the doubled frequencies.
the dash-dotted magenta lines show sums of frequencies. The insets show
the corresponding spectrum  of the visibility, $\mathcal F(\omega_n)$ in Eq.~\eqref{F:0}. The parameter $g$ is, from left to right, $0.02,~ -0.005$, and $-0.1$, while the critical value is $g_c=-0.0165$, and $\Delta=0.025$.}
\label{fig:spectrum}
\end{figure*}

Further information is gained by inspecting the Fourier transform of the visibility and of its logarithm, respectively defined as
\begin{align}
\label{F:0}
 \mathcal F(\omega_n) &= \frac{1}{T} \int_0^T \mathrm dt \; \mathcal V(t) \; e^{-{\mathrm{i}} \omega_n t}\,,\\
\label{F:log}
 \mathcal F_L(\omega_n) &= \frac{1}{T} \int_0^T \mathrm dt \; \ln [ \mathcal V(t) ] \; e^{-{\mathrm{i}} \omega_n t}       \,,
\end{align}
where $\omega_n=2\pi n/T$ with $n \in \mathbb{N}_0$ and $T$ is a time interval such that $T\nu_{\rm min}\gg1$, with $\hbar\nu_{\rm min}$ the smallest gap in Hamiltonian $H_{\rm eff}^{(e)}$. Figure~\ref{fig:spectrum} displays the spectra corresponding to the visibility as a function of time in Fig.~\ref{fig:signals}, and its logarithm. For $g>0$, shown in (a), the main peak is located at twice the eigenfrequency of the zigzag mode, while for $g<0$ it is at the eigenfrequency of the corresponding lowest frequency mode, see (b) and (c), which becomes unstable when the critical value is approached. The behaviour for $g>0$ hints to the presence of squeezing, originated by quenching the trap frequency (and thus the normal mode frequencies) by exciting the central ion. The frequency of oscillations of the visibility observed in Fig.~\ref{fig:signals}(a) corresponds indeed to $2\omega_1^e$, which is twice the frequency of the zigzag mode of the linear chain. For $g<0$, the main peaks are also associated with the vibrational mode that becomes unstable at the critical point, and which is the one that is most significantly excited by the quench. The main peak is now at $\omega_1^e$ instead of $2\omega_1^e$ because the dominant effect of the quench is the displacement of the equilibrium positions. In (b) and (c) minor peaks are present at the eigenfrequencies of the modes which couple to the zigzag mode and in (b) also at sums of eigenfrequencies.

\begin{figure}[htbp]
  \centering
  \includegraphics[width=0.45\textwidth]{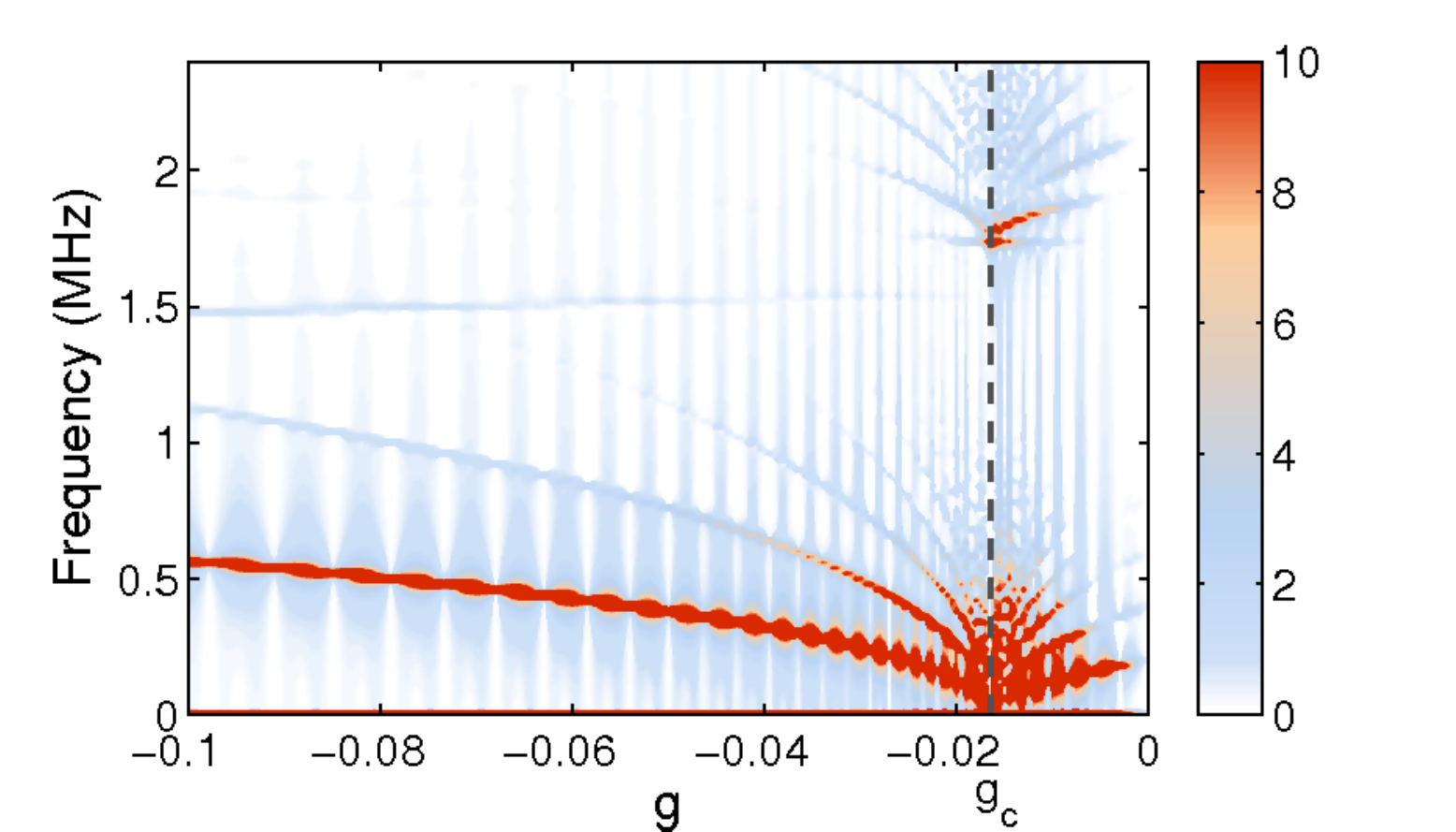}
  \caption{(Color online) Density plot of the logarithmic spectrum $\mathcal F_L(\omega)$ of the visibility for the parameters of Fig.~\ref{fig:curv}, $\Delta = 0.025$, and $g<0$. The dashed line shows the critical value $g_c=-0.0165$.}
  \label{fig:spectrumcontour}
\end{figure}

The spectrum of the logarithmic signal as a function of $g$ is displayed in Fig.~\ref{fig:spectrumcontour} for $g<0$ and $\Delta$ constant. One observes a main peak corresponding to the mode whose frequency goes to zero when $g$ approaches $g_c$ and the crystal in the excited state becomes unstable. Close to $g_c$ one observes two signals which become more visible, that are at twice the zigzag eigenfrequency and at the sum of the zigzag mode frequency with the axial breathing mode frequency. They hint at the presence of single-mode and multimode squeezing due to the quenching, suggesting that some entanglement between the modes is generated by the quench. 

We now summarize our findings. In the first place, the visibility decays fast to zero when the quench is performed between motional states whose classical equilibrium configurations differ. The decay is faster when the two structures have different symmetries, and thus different spectral properties, such as when the quench is across the phase transition linear-zigzag. Nevertheless, the visibility exhibits revivals as a function of the time after the quench. These revivals occur at the frequency of the lowest normal mode of the zigzag structure, which becomes unstable at the instability and corresponds to the zigzag mode of the linear chain. This mode is in fact the one which has the largest overlap with the difference between zigzag and linear structures. This feature of the visibility is scalable, the periodicity of the revival remains in fact finite as the size of the chain is increased. The height of the peaks, however, decays as $N$ is scaled up, consistently with the fact that the amplitude of excitation of the zigzag mode by a displacement of the central ion decreases as the size of the chain is increased. In the thermodynamic limit, hence, the surviving feature is the decay of the visibility signal at short times, corresponding to the fact that the quantum superposition of the spin irreversibly dephases. When instead the quench is between two linear chains, there is a quasi-periodic rephasing of the system, corresponding to the creation of squeezing by sudden changing the trap frequency and thus the normal mode frequencies~\cite{Heinzen}. The amplitude of the oscillations decays to zero while the visibility reaches a constant value which approaches unity as $g$ is increased.
 
 Before we conclude this subsection, we remark that the dynamics we consider in this paper is unitary: decay of the overlap signal is solely due to dephasing in the dynamics of internal and external degrees of freedom, while external sources of decoherence and noise have been neglected. They are expected to introduce a damping factor in the overlap signal, setting an upper bound for partial revivals decreasing with time. Some estimates have been already provided in Ref.~\cite{Baltrusch2011}, where it was argued that for the parameters of existing set-ups the revivals of the visibility should be observed. In particular, the main source of noise in ion traps is considered to be patch potentials at the electrodes~\cite{Roos,Porras}. The measured heating rates depend strongly on the actual set-up of the experimental apparatus, the smaller heating rates which have been reported correspond to timescales of the order of several milliseconds, which would allow one to observe several revivals of the visibility. The other important point regards the assumption that the chain is initially in the ground state of the vibrational excitations.  Ground state cooling of ion chains composed to up to 4 ions have been successfully demonstrated in~\cite{NIST_Jost}, The visibility signals is degraded as the temperature $T$ of the chain is increased. The functional dependence of the visibility on $T$ is subject of ongoing studies.

\section{Conclusions}
\label{sec:conclusions}

The dynamical properties of an ion crystal after a quench have been theoretically investigated, when the quench is performed by creating coherent superpositions of motional states close to and across the linear-zigzag structural transition. These dynamics have been related to the visibility of the signal when Ramsey interferometry is performed on one ion of the chain. The visibility decays at short times as the internal state becomes entangled with the motional state of the crystal, but exhibits periodic revivals at longer times, determined by the frequency of the zigzag mode. Further periodic signals appear at multiples of the zigzag mode and at sums of different motional excitations, suggesting squeezing and entanglement in the vibrational motion generated by the quench of the trap frequency. These spectral properties persist as the number of ions increases, even though the heights of the revivals decrease. These results are based on a theoretical model which we report in detail and which allows one to calculate the visibility for different parameter regimes. This model is valid as long as as the harmonic theory of the crystal is applicable and is thus reliable for the parameters we consider in this paper. 

Our analysis shows that, if the crystal is initially in the motional ground state, these features can be observed for  parameters that are consistent with ongoing experimental work. The signal is however degraded as the temperature is increased, the functional dependence of the visibility signal on the chain temperature is object of ongoing studies. 
 
We conclude by observing that the visibility signal allows one to study the behaviour of the soft mode across the classical phase transition. Extensions of these studies to the parameter regime where quantum effects at the phase transition are relevant~\cite{Retzker,Shimshoni} would allow one to extract the corresponding quantum fidelity, in the spirit of the work performed in~\cite{Qfidelity}, and will be subject of future studies.

\section*{Acknowledgments}
The authors acknowledge discussions with Tommaso Calarco, Gabriele De Chiara, and Shmuel Fishman and support by the European Commission (Integrating Project ``AQUTE'', STREP ``PICC'', COST action ``IOTA''), the Spanish Ministry of Science (EUROQUAM ``CMMC'', Consolider Ingenio 2010), the Alexander von Humboldt and the German Research Foundations.  

\appendix
\section{Multimode Squeezing Operator and disentanglement theorem}
  \label{sec:disentangling}
  
In order to obtain Eq.~\eqref{eq:Squeez:1} from Eq.~\eqref{eq:Squeez} we follow the method of Bogoliubov and Shirkov~\cite{BogoliubovShirkov}. The basic idea is best understood by first considering operator $e^{\lambda(\sigma_+ + \sigma_-)} $, with $\lambda$ scalar, $\sigma_z$ the Pauli matrix and $\sigma_\pm$ the raising and lowering operators for a spin 1/2, such that $ \commutator{\sigma_+}{\sigma_-} = \sigma_z $, $ \commutator{\sigma_z}{\sigma_\pm} = \pm 2 \sigma_\pm $. The disentangling formula reads:
\begin{equation}
\label{eq:disentangle}
e^{\lambda(\sigma_+ + \sigma_-)} = e^{ \sigma_+ \tanh(\lambda) } e^{-\sigma_z \ln \left[ \cosh(\lambda) \right]} e^{ \sigma_- \tanh(\lambda) } \,,
\end{equation}
and can be obtained using the procedure sketched in Ref.~\cite{ColletPRA1988}. Assuming $ \lambda $ to be a continuous parameter, one makes the ansatz  $ e^{\lambda (\sigma_+ + \sigma_-)} = e^{f(\lambda) \sigma_+} F(\lambda) e^{g(\lambda)\sigma_-}$ with $F(\lambda)$ an operator such that $F(0) = 1$, while $f(\lambda)$ and $g(\lambda)$ are analytic functions of $\lambda$ with $f(0)=0$ and $g(0)=0$. Under these assumptions $ F(\lambda) $ can be written as
\begin{equation}
F(\lambda) = e^{-f(\lambda)\sigma_+} e^{\lambda(\sigma_+ + \sigma_-)} e^{-g(\lambda) \sigma_-} \,.
\end{equation}
We take the derivative of $F(\lambda)$ and obtain a first-order differential equation which contains all operators. The contributions from $ \sigma_+ $ and $ \sigma_- $ cancel out by choosing $f(\lambda) = g(\lambda)= \tanh (\lambda)$, so that $F(\lambda) = \exp \left[ - \sigma_z \ln \left( \cosh \lambda \right) \right]$, hence demonstrating Eq.~\eqref{eq:disentangle}.

This procedure can be generalized to show the equality
\begin{equation}
e^{\frac{\lambda}{2}(a^\dagger{}^2 - a^2)} = 
e^{\frac{\tanh\lambda}{2}a^\dagger{}^2} e^{- (a^\dagger a + \frac{1}{2} ) \ln\left[\cosh\lambda\right]} e^{-\frac{\tanh\lambda}{2}a^2}
\end{equation}
where $a, a^\dagger$ are the annihilation and creation operators of a harmonic oscillator.

Moreover, we can use the procedure sketched above in order to disentangle the multimode squeezing operator:
\begin{multline}
 \label{eq:disentangling}
 \exp \biggl\lbrace \frac{1}{2}\sum_{jk} \left( \xi_{jk} \creation{a_j}  \creation{a_k} - \xi^*_{jk} \annihilation{a_j} \annihilation{a_k} \right) \biggr\rbrace \\
= Z \; e^{\frac{1}{2}\sum_{jk} A_{jk}  a_j{\!}^\dagger a_k{\!}^\dagger} 
		e^{- \sum_{jk} B_{jk}  a_j{\!}^\dagger a_k } \;
		e^{ -\frac{1}{2} \sum_{jk} C_{jk} \, a_j a_k } \,,
\end{multline}
with
\begin{align}
A_{jk} &= \sum_l \tanh(\chi_l) \Lambda_{jl} \Lambda_{kl} \,,\\
B_{jk} &= \sum_l \ln\bigl(\cosh\chi_l\bigr) \Lambda_{jl} \Lambda_{kl}^* \,,\\
C_{jk} &= \sum_l \tanh(\chi_l) \Lambda_{jl}^* \Lambda_{kl}^* = A_{jk}^* \label{eq:disentangling:Amatrix}\,,\\
Z_{\phantom{jk}} &= \exp \biggl \lbrace - \sum_j \frac{1}{2} \ln\bigl(\cosh\chi_j\bigr) \biggr \rbrace \,.
\end{align}
This can be done after observing that, since $ \xi $ is complex symmetric, we can perform Takagi's factorization~\cite{HornJohnson} $ \xi = \Lambda \chi \Lambda^T $, where $ \Lambda $ is unitary and $ \chi = \diag(\lbrace \chi_1, \chi_2, \dotsc \rbrace)$ is diagonal with $ \chi_j \geq 0$ real and non-negative (this factorization exists for any complex symmetric matrix). This defines the transformation $ b_j{}^\dagger = \sum_{jk}\Lambda_{kj} a_k{}^\dagger $, $ b_j = \sum_{jk}\Lambda_{kj}^* a_k $ for a new set of operators for which the squeezing operator is in diagonal form, $ \exp \left[ \frac{1}{2}\sum_{j} \chi_{j} \left ( b_j{}^\dagger{}^2 - b_j{}^2 \right) \right] $. These new operators have bosonic commutation relations, $ \commutator{b_j}{b_k} = \commutator{b_j{}^\dagger}{b_k{^\dagger}} = 0 $, and $ \commutator
 {b_j}{b_k{}^\dagger} = \sum_l \Lambda_{lk} \Lambda_{lj}^* = \delta_{jk} $ since $ \Lambda $ is unitary.
Therefore, operators of different modes factorize as $\prod_j \exp \left\lbrace \frac{\chi_j}{2} \left( b_j{}^\dagger{}^2 - b_j{}^2 \right) \right\rbrace$, and one finally obtains
\begin{multline*}
 \exp \biggl\lbrace \frac{1}{2}\sum_{jk} \left( \xi_{jk} \creation{a_j}  \creation{a_k} - \xi^*_{jk} \annihilation{a_j} \annihilation{a_k} \right) \biggr\rbrace \\
= \prod_j e^{\frac{\tanh\chi_j}{2}b_j{}^\dagger{}^2} e^{- (b_j{}^\dagger b_j + \frac{1}{2} ) \ln\bigl(\cosh\chi_j\bigr)} e^{-\frac{\tanh\chi_j}{2}b_j{}^2} \,.
\end{multline*}
The terms belonging to different modes commute now, so bringing factors with operators $b^{\dagger 2}_j$ to the left and factors with $b_j^2$ to the right and writing them as a function of operators $a_k$ and $a_k^{\dagger}$, one obtains Eq.~\eqref{eq:disentangling}.

\section{Calculation of the Normalization Constant}
 \label{sec:linkedcluster}

In order to calculate the constant $Z$ we use the normalization condition of the states as stated in Eq.~\eqref{eq:normalization}, 
\begin{align*}
1  &= Z^2 \matrixElement{0}{ \left( \sum_{n=0}^\infty \sum_{m=0}^\infty \frac{{\mathrm{A}^\dagger}^n {\mathrm{A}^{\vphantom \dagger}}^m}{n! m!} \right)}{0} \,.
\end{align*}
Since $A$ contains only creation operators, only the summands with $m=n$ give a contribution:
\begin{align}
Z^{-2} &= \matrixElement{0}{\left( \sum_{n=0}^\infty \frac{{\mathrm{A}^\dagger}^n {\mathrm{A}^{\vphantom \dagger}}^n}{(n!)^2} \right)}{0} 	\nonumber\\
&= \sum_{n=0}^\infty \frac{1}{( n!)^2} \matrixElement {0}{{\mathrm{A}^\dagger}^n {\mathrm{A}^{\vphantom \dagger}}^n }{0} = \sum_{n=0}^\infty W_n,
\label{eq:normalization1}
\end{align}
where $W_n$ is defined as 
\begin{multline}
\label{def:Wn}
W_n = \frac{1}{(2^n n!)^2}\sum_{\substack{j_1 \dotsm j_{2n} \\ k_1 \dotsm k_{2n}}} A_{j_1 j_2} \dotsm A_{j_{2n-1} j_{2n}} A_{k_1 k_2} \dotsm \\ 
\dotsm A_{k_{2n-1}k_{2n}} \matrixElement{0}{b_{j_1}^{\vphantom{\dagger}} \dotsm b_{j_{2n}}^{\vphantom{\dagger}} b_{k_1}^\dagger \dotsm b_{k_{2n}}^\dagger 
}{0}.
\end{multline}
The sum contains $(2n)!$ summands (which contain $2n$ Kronecker-delta symbols) which do not vanish, corresponding to the number of all pairs of sets of indices $\{j_1 \dotsm j_{2n}\}$ and $\{k_1 \dotsm k_{2n}\}$ which are identical (apart for a permutation within the same set). For example, 
\begin{equation*}
W_1 = \frac{1}{2^2} \sum_{\substack{j_1 j_2\\ k_1 k_2}} A_{j_1 j_2}A_{k_1 k_2} \Big\{ \delta_{j_1 k_1} \delta_{j_2 k_2} + \delta_{j_1 k_2} \delta_{j_2 k_1} \Big\}.
\end{equation*}
while $W_2$ has already 24 summands, we write only two of them exemplarily:
\begin{align}
\sum_{\substack{j_1 j_2 j_3 j_4\\ k_1 k_2 k_3 k_4}} A_{j_1 j_2}A_{j_3 j_4}A_{k_1 k_2}A_{k_3 k_4} \cdot \delta_{j_1 k_4} \delta_{j_2 k_1} \delta_{j_3 k_2} \delta_{j_4 k_1} \label{eq:graph1}\\
\sum_{\substack{j_1 j_2 j_3 j_4\\ k_1 k_2 k_3 k_4}} A_{j_1 j_2}A_{j_3 j_4}A_{k_1 k_2}A_{k_3 k_4} \cdot \delta_{j_1 k_1} \delta_{j_2 k_2} \delta_{j_3 k_4} \delta_{j_4 k_3} \label{eq:graph2}
\end{align}
We now associate with each summand in $W_n$ a graph, which we call a \emph{$n$-graph}. For this, let the first $n$ coefficients be represented by $n$ pairs of adjacent circles in an upper row, while the second $n$ coefficients are represented by the same number of pairs of circles in the lower row. The indices $\{j_1 \dotsm j_{2n}\}$ and $\{k_1 \dotsm k_{2n}\}$ are filled in in correct ordering into the circles such that there are only $j$'s in the upper and only $k$'s in the lower row.
\begin{center}
\begin{tikzpicture}
\tikzstyle{every state}=[fill=none,draw=black,text=black,inner sep=0pt,minimum size=4mm,		node distance=0mm]
  \node[black]			at (  -1,1.2  )		 {$j$:};	
  \node[black]  		at (  -1,0  )		 {$k$:};	
  \node[state]	(j1)	at (0  ,1.2)     {\scriptsize 1};
  \node[state]	(j2)	at ( .4,1.2)     {\scriptsize 2};
  \node[state]	(j3)	at (1.2,1.2)     {\scriptsize 3};
  \node[state]	(j4)	at (1.6,1.2)     {\scriptsize 4};
  \node[black]			at (  2.4,1.2  )		 {$\dotsm$};	
  \node[black]  		at (  2.4,0  )		 {$\dotsm$};
  \node[state]	(k1)	at (  0,0  )     {\scriptsize 1};
  \node[state]	(k2)	at ( .4,0  )     {\scriptsize 2};
  \node[state]	(k3)	at (1.2,0  )     {\scriptsize 3};
  \node[state]	(k4)	at (1.6,0  )     {\scriptsize 4};
  \draw (-1.6, 1.6) to (-1.6,- .4);
  \draw (-1.6, 1.6) to (-1.4, 1.6);
  \draw (-1.6,- .4) to (-1.4,- .4);
  \draw ( 3  , 1.6) to ( 3  ,- .4);
  \draw ( 2.8, 1.6) to ( 3  , 1.6);
  \draw ( 2.8,- .4) to ( 3  ,- .4);
\end{tikzpicture}
\end{center}
Then for each Kronecker-$\delta$ we need to connect the corresponding two circles by a line. We find easily that each circle must be connected with another, and that there is a total of $2n$ lines. Thus each circle has exactly \emph{one} line. For instance, the graphs for~\eqref{eq:graph1} and~\eqref{eq:graph2} are given by: 
\begin{center}
\begin{tikzpicture}
\tikzstyle{every state}=[fill=none,draw=black,text=black,inner sep=0pt,minimum size=4mm,		node distance=0mm]

  \node[state]	(j1)	at (0  ,1.2)     {\scriptsize 1};
  \node[state]	(j2)	at ( .4,1.2)     {\scriptsize 2};
  \node[state]	(j3)	at (1.2,1.2)     {\scriptsize 3};
  \node[state]	(j4)	at (1.6,1.2)     {\scriptsize 4};
  \node[state]	(k1)	at (  0,0  )     {\scriptsize 1};
  \node[state]	(k2)	at ( .4,0  )     {\scriptsize 2};
  \node[state]	(k3)	at (1.2,0  )     {\scriptsize 3};
  \node[state]	(k4)	at (1.6,0  )     {\scriptsize 4};
  \draw (0  ,1  ) to (1.6, .2);
  \draw (0.4,1  ) to (  0, .2);
  \draw (1.2,1  ) to ( .4, .2);
  \draw (1.6,1  ) to (1.2, .2);
  \draw (- .6, 1.6) to (- .6,- .4);
  \draw (- .6, 1.6) to (- .4, 1.6);
  \draw (- .6,- .4) to (- .4,- .4);
  \draw ( 2.2, 1.6) to ( 2.2,- .4);
  \draw ( 2  , 1.6) to ( 2.2, 1.6);
  \draw ( 2  ,- .4) to ( 2.2,- .4);
\end{tikzpicture}
\hspace{5pt}
\raisebox{25pt}{and}
\hspace{5pt}
\begin{tikzpicture}
\tikzstyle{every state}=[fill=none,draw=black,text=black,inner sep=0pt,minimum size=4mm,		node distance=0mm]

  \node[state]	(j1)	at (0  ,1.2)     {\scriptsize 1};
  \node[state]	(j2)	at ( .4,1.2)     {\scriptsize 2};
  \node[state]	(j3)	at (1.2,1.2)     {\scriptsize 3};
  \node[state]	(j4)	at (1.6,1.2)     {\scriptsize 4};
  \node[state]	(k1)	at (  0,0  )     {\scriptsize 1};
  \node[state]	(k2)	at ( .4,0  )     {\scriptsize 2};
  \node[state]	(k3)	at (1.2,0  )     {\scriptsize 3};
  \node[state]	(k4)	at (1.6,0  )     {\scriptsize 4};
  \draw (0  ,1  ) to (  0, .2);
  \draw (0.4,1  ) to ( .4, .2);
  \draw (1.2,1  ) to (1.6, .2);
  \draw (1.6,1  ) to (1.2, .2);
  \draw (- .6, 1.6) to (- .6,- .4);
  \draw (- .6, 1.6) to (- .4, 1.6);
  \draw (- .6,- .4) to (- .4,- .4);
  \draw ( 2.2, 1.6) to ( 2.2,- .4);
  \draw ( 2  , 1.6) to ( 2.2, 1.6);
  \draw ( 2  ,- .4) to ( 2.2,- .4);
\end{tikzpicture}
\end{center}
respectively. If we evaluate~\eqref{eq:graph1}, we find that it yields $\trace (\mathrm{A}^4)$, while~\eqref{eq:graph2} can be factorized into two terms 
\begin{equation}
\left[\sum_{\substack{j_1 j_2\\ k_1 k_2}} A_{j_1 j_2}A_{k_1 k_2}\delta_{j_1 k_1} \delta_{j_2 k_2}\right]\cdot\left[\sum_{\substack{j_3 j_4\\k_3 k_4}}A_{j_3 j_4}A_{k_3 k_4}  \delta_{j_3 k_4} \delta_{j_4 k_3} \right] \nonumber ,
\end{equation}
which give $[\trace (\mathrm{A}^2)]^2$. This factorization can also be shown graphically, 
\begin{center}
\begin{tikzpicture}
\tikzstyle{every state}=[fill=none,draw=black,text=black,inner sep=0pt,minimum size=4mm,		node distance=0mm]

  \node[state]	(j1)	at (0  ,1.2)     {\scriptsize 1};
  \node[state]	(j2)	at ( .4,1.2)     {\scriptsize 2};
  \node[state]	(j3)	at (1.2,1.2)     {\scriptsize 3};
  \node[state]	(j4)	at (1.6,1.2)     {\scriptsize 4};
  \node[state]	(k1)	at (  0,0  )     {\scriptsize 1};
  \node[state]	(k2)	at ( .4,0  )     {\scriptsize 2};
  \node[state]	(k3)	at (1.2,0  )     {\scriptsize 3};
  \node[state]	(k4)	at (1.6,0  )     {\scriptsize 4};
  \draw (0  ,1  ) to (  0, .2);
  \draw (0.4,1  ) to ( .4, .2);
  \draw (1.2,1  ) to (1.6, .2);
  \draw (1.6,1  ) to (1.2, .2);
  \draw (- .6, 1.6) to (- .6,- .4);
  \draw (- .6, 1.6) to (- .4, 1.6);
  \draw (- .6,- .4) to (- .4,- .4);
  \draw ( 2.2, 1.6) to ( 2.2,- .4);
  \draw ( 2  , 1.6) to ( 2.2, 1.6);
  \draw ( 2  ,- .4) to ( 2.2,- .4);
\end{tikzpicture}
\raisebox{25pt}{$\equiv$}
\begin{tikzpicture}
\tikzstyle{every state}=[fill=none,draw=black,text=black,inner sep=0pt,minimum size=4mm,		node distance=0mm]
  \draw (- .6, 1.6) to (- .6,- .4);
  \draw (- .6, 1.6) to (- .4, 1.6);
  \draw (- .6,- .4) to (- .4,- .4);
  \node[state]	(j1)	at (0  ,1.2)     {\scriptsize 1};
  \node[state]	(j2)	at ( .4,1.2)     {\scriptsize 2};
  \node[state]	(k1)	at (  0,0  )     {\scriptsize 1};
  \node[state]	(k2)	at ( .4,0  )     {\scriptsize 2};
  \draw (0  ,1  ) to (  0, .2);
  \draw (0.4,1  ) to ( .4, .2);
  \draw ( 1  , 1.6) to ( 1  ,- .4);
  \draw (  .8, 1.6) to ( 1  , 1.6);
  \draw (  .8,- .4) to ( 1  ,- .4);
  \node [circle,fill=black,inner sep=.5pt] at ( 1.3,0.6) {}; 
  \draw ( 1.6, 1.6) to ( 1.6,- .4);
  \draw ( 1.6, 1.6) to ( 1.8, 1.6);
  \draw ( 1.6,- .4) to ( 1.8,- .4);
  \node[state]	(j3)	at (2.2,1.2)     {\scriptsize 3};
  \node[state]	(j4)	at (2.6,1.2)     {\scriptsize 4};
  \node[state]	(k3)	at (2.2,0  )     {\scriptsize 3};
  \node[state]	(k4)	at (2.6,0  )     {\scriptsize 4};
  \draw (2.2,1  ) to (2.6, .2);
  \draw (2.6,1  ) to (2.2, .2);
  \draw ( 3.2, 1.6) to ( 3.2,- .4);
  \draw ( 3  , 1.6) to ( 3.2, 1.6);
  \draw ( 3  ,- .4) to ( 3.2,- .4);
\end{tikzpicture}.
\end{center}

Thus, a graph may be decomposed into a product of fully connected subgraphs or clusters. An $n$-graph can be decomposed into a product of $m_1$ 1-clusters, $m_2$ 2-clusters, $\dotsc$, and $m_n$ $n$-clusters, where the $m_l$ fulfill 
\begin{equation}
\sum_{l=1}^n m_l \,l = n.\label{eq:restrictionml}
\end{equation}
The evaluation of an $l$-cluster always yields $\trace (\mathrm{A}^{2l})$, and there are $2^l \, l!\; 2^{l-1} (l-1)!$ ways to draw an $l$-cluster. So we are motivated to define the \emph{$l$-cluster integral} by the sum of all possible clusters for $l$ pairs of circles in each row, which after evaluation is given by:
\begin{equation}
b_l = 2^l \, l!\; 2^{l-1} (l-1)! \trace (\mathrm{A}^{2l}).\label{eq:lcluster}
\end{equation}
We have $b_0=1$ and $b_1= 2 \trace (\mathrm{A}^2)$, which finds its graphical representation by 
\begin{center}
\begin{tikzpicture}
\tikzstyle{every state}=[fill=none,draw=black,text=black,inner sep=0pt,minimum size=4mm,		node distance=0mm]
  \draw (- .6, 1.6) to (- .6,- .4);
  \draw (- .6, 1.6) to (- .4, 1.6);
  \draw (- .6,- .4) to (- .4,- .4);
  \node[state]	(j1)	at (0  ,1.2)     {};
  \node[state]	(j2)	at ( .4,1.2)     {};
  \node[state]	(k1)	at (  0,0  )     {};
  \node[state]	(k2)	at ( .4,0  )     {};
  \draw (0  ,1  ) to (  0, .2);
  \draw (0.4,1  ) to ( .4, .2);
  \node  at ( 1.1,0.6) {+}; 
  \node[state]	(j3)	at (1.8,1.2)     {};
  \node[state]	(j4)	at (2.2,1.2)     {};
  \node[state]	(k3)	at (1.8,0  )     {};
  \node[state]	(k4)	at (2.2,0  )     {};
  \draw (1.8,1  ) to (2.2, .2);
  \draw (2.2,1  ) to (1.8, .2);
  \draw ( 2.8, 1.6) to ( 2.8,- .4);
  \draw ( 2.6, 1.6) to ( 2.8, 1.6);
  \draw ( 2.6,- .4) to ( 2.8,- .4);
\end{tikzpicture}.
\end{center}
Accordingly one can draw the cluster-integrals for the higher orders. 
Here we have not filled out the circles, since the cluster integral is independent of the indices which are assigned to it. It is clear that for a given set of indices $\{i_1,i_2,\dotsc\}$ the circles have to be filled in the same ordering in each summand, and without loss of generality one can fill the circles in the natural ordering $(j_{i_1},j_{i_2},\dotsc)$ and $(k_{i_1},k_{i_2},\dotsc)$ where $i_1<i_2<\dotsb$.
The total set of indices cannot be split arbitrarily in between the clusters, since pairs of indices of the form $(j_{2l-1},j_{2l})$ always belong to the same cluster.

To resume the calculation, we note that
\begin{equation} \label{eq:Wsum}
W_n = \frac{1}{(2^n n!)^2} {\sum_{\{m_l\}}}^\prime \mathcal S\{m_l\},
\end{equation}
where $\mathcal S\{m_l\}$ is the sum over all possible graphs described by the set of integers $\{m_l\}$, and the primed sum denotes a restricted summation over all sets $\{m_l\}$ which fulfill equation~\eqref{eq:restrictionml}.
We see that 
\begin{equation} \label{eq:sumoverallgraphsml}
\mathcal S\{m_l\} = \sum_{\mathcal P} b_1^{m_1} b_2^{m_2} \dotsm , 
\end{equation}
where the summation over $\mathcal P$ extends over all possible ways of distributing the two times $n$ pairs of indices $\{(j_1;j_2),\dotsc,(j_{2n-1};j_{2n})\}$ and $\{(k_1;k_2),\dotsc,(k_{2n-1};k_{2n})\}$ into the circles obtaining only distinct graphs. So there are $n!$ ways of distributing these pairs (the ordering of a pair is already contained inside the cluster integral). A permutation of two $l$-clusters with the same $l$ does not give a new graph, therefore we get a factor $\prod_l (m_l !)^{-1}$. Moreover, a permutation of pairs inside a cluster integral does not give a new graph either. Thus we get a factor $\prod_l (l!)^{-2m_l}$. Equation~\eqref{eq:sumoverallgraphsml} is then given by
\begin{equation}
\mathcal S\{m_l\} = (n!)^2 \prod_{l=1}^n \frac{b_l^{m_l}}{m_l! (l!)^{2m_l}}.
\end{equation}
Replacing in (\ref{eq:Wsum}) one gets:
\begin{align}
W_n &= \frac{1}{(2^n n!)^2} {\sum_{\{m_l\}}}^\prime (n!)^2 \prod_{l=1}^n \frac{b_l^{m_l}}{m_l! (l!)^{2m_l}} \nonumber \\
	&= \frac{1}{2^{2n}} {\sum_{\{m_l\}}}^\prime \prod_{l=1}^n \frac{1}{m_l!} \left(\frac{b_l}{(l!)^2} \right)^{m_l} \nonumber\\
	&=  {\sum_{\{m_l\}}}^\prime \frac{1}{2^{2(m_1 1 + m_2 2 + \dotsb)  }} \prod_{l=1}^n \frac{1}{m_l!} \left(\frac{b_l}{(l!)^2} \right)^{m_l} \nonumber\\
	&=  {\sum_{\{m_l\}}}^\prime \prod_{l=1}^n \frac{1}{m_l!} \left(\frac{b_l}{(2^l l!)^2} \right)^{m_l}.
\end{align}
We can now insert this result in Eq. (\ref{eq:normalization1}) and obtain:
\begin{equation}
Z^{-2} = \sum_{n=0}^\infty {\sum_{m_l}}^\prime \prod_{l=1}^n \frac{1}{m_l!} \left(\frac{b_l}{(2^l l!)^2} \right)^{m_l} .
\end{equation}
Summing over all $\{m_l\}$ followed by summation over all $n$ is equivalent to summing over all $m_1, m_2,\dotsc$ from $0$ to $\infty$ separately, so we can replace the restricted sum:
\begin{align}
Z^{-2}	&= \sum_{m_1=0}^\infty \sum_{m_2=0}^\infty \dotsb \prod_{l=1}^\infty \frac{1}{m_l!} \left(\frac{b_l}{(2^l l!)^2} \right)^{m_l} \nonumber\\
		&= \prod_{l=1}^\infty \left[ \sum_{m_l=0}^\infty \frac{1}{m_l!} \left(\frac{b_l}{(2^l l!)^2}\right)^{m_l} \right] \nonumber\\
		&= \prod_{l=1}^\infty \exp{\left[ \frac{b_l}{(2^l l!)^2} \right]} =  \exp{\left[ \sum_{l=1}^\infty \frac{b_l}{(2^l l!)^2} \right]} \,.
\end{align}
Using equation~\eqref{eq:lcluster}, we finally get
\begin{align}
Z^{-2}= \exp{\left[ \sum_{l=1}^\infty \frac{\trace (\mathrm{A}^{2l})}{2l} \right]}= \exp{\Bigg( \frac{1}{2} \trace \bigg[ \ln{\frac{1}{1-\mathrm{A}^2} }\bigg]\Bigg) } \,,
\label{Z:A}
\end{align}
which is valid if $ 1-  \mathrm{A}^2 $ is non-singular. To show that this is true it is sufficient to show that any matrix norm of $ \mathrm{A} $ is smaller than one. Using the spectral norm $\matrixNorm \cdot$, the form of Eq.~\eqref{eq:disentangling:Amatrix}, and the submultiplicativity of the matrix norm, we have $ \matrixNorm{\mathrm{A}}_2 \leq \matrixNorm{\Lambda}_2 \matrixNorm{\tanh \chi}_2 \matrixNorm{\Lambda^T}_2 = \matrixNorm{\tanh \chi}_2$. Using the fact that $ \chi $ is diagonal, real and positive, the spectral norm is equal to tangent hyperbolicus of the largest eigenvalue of $ \chi $. Thus $ \matrixNorm{\mathrm{A}}_2 < 1 $ as the tangent hyperbolicus is smaller than 1 in its full domain. Equation~\eqref{Z:A} can thus be cast in the compact form:
\begin{align}
Z = \exp \left( -\frac{1}{4} \trace \left[  \ln \frac{1}{1-\mathrm{A}^2} \right] \right)= \det \left[ \left( 1-\mathrm{A}^2 \right)^{\frac{1}{4}} \right] \,.
\end{align}
\section{Calculation of the Overlap Integral}
  \label{sec:overlapintegral}
We consider the integral~\eqref{eq:O:3} and first remove the time-dependent phase factors $ e^{-{\mathrm{i}} \omega_j t} $ from the integration variables $ \alpha_j $ in $\mathit{a}^*(\alpha(t) -  \beta)$ by shifting it to the coefficients $A_{jk}$ by defining $A_{jk}(t) =  A_{jk} e^{-{\mathrm{i}} (\omega_j+\omega_k) t}$. We merge all terms into a single exponential whose exponent reads
 \begin{multline*}
  \frac{1}{2} \sum_{jk} 
  \begin{pmatrix}
    \alpha_j \\ \alpha_j^*
    \end{pmatrix}^{T}
  \begin{pmatrix}
    A_{jk}(t) & -\delta_{jk} \\ -\delta_{jk} & A_{jk} 
  \end{pmatrix}
  \begin{pmatrix}
    \alpha_k \\ \alpha_k^*
  \end{pmatrix} \\
 - \sum_j S_{j}[\beta^*] \alpha_j^* - \sum_j S_{j}[\beta] e^{-{\mathrm{i}}\omega_j t} \alpha_j +  G^*(\beta) + G(\beta) \,,
\end{multline*}
with 
\begin{equation}
  S_j[\beta] = \sum_{k} A_{jk} \beta_k - \beta_j^* \,,
\end{equation}
and
\begin{equation}
  G(\beta) =   \sum_{jk}  \frac{A_{jk}}{2} \beta_j^*\beta_k^*   - \sum_{j} \frac{|\beta_j|^2}{2} \,.
\end{equation}
We now express the integration variables by their real and imaginary parts, $\alpha_j = u_j + {\mathrm{i}} v_j$ and $\alpha_j^* = u_j - {\mathrm{i}} v_j$. 
The quadratic term is written as
\begin{equation*}
 - \sum_{jk}
 \begin{pmatrix}
  u_j \\ v_j 
 \end{pmatrix}^T
\begin{pmatrix}
 \delta_{jk} - \Lambda_{jk}^+ & \phantom{\delta_{jk}}-{\mathrm{i}} \Lambda_{jk}^- \\ \phantom{\delta_{jk}}-{\mathrm{i}} \Lambda_{jk}^- & \delta_{jk}+ \Lambda_{jk}^+
\end{pmatrix}
\begin{pmatrix}
 u_k \\ v_k
\end{pmatrix}
\end{equation*}
with complex symmetric matrices 
\begin{equation}
\Lambda_{jk}^\pm =  \frac{1}{2} \bigl ( A_{jk}(t) \pm A_{jk}(0) \bigr )   \,.
\end{equation}
The linear term in the exponent takes the form $ - \sum_j \left[S^+_j u_j - {\mathrm{i}} S^-_j v_j\right]$  with 
\begin{equation}
 S^\pm_j = S_{j}[\beta^*] \pm S_{j}[\beta] e^{-{\mathrm{i}}\omega_j t}\,.
\end{equation}
Introducing the vector $\vek w = (u, v)^T$ where $u = (u_1,\dotsc ,u_{2N})$, $v = (v_1,\dotsc ,v_{2N})$, we can write the overlap as
\begin{equation}
\label{eq:O:4}
 \mathcal O(t) = \frac{Z^2 }{\pi^{2N}} e^{G^*(\beta)}e^{G(\beta)} \int \mathrm{d} \vek w e^{ -\vek w^T . \vek w  - \vek w^T \Omega \vek w } \,,
\end{equation}
with 
\begin{align}
 \vek s &= \begin{pmatrix}
           \phantom{-{\mathrm{i}}}S^+ \\ -{\mathrm{i}} S^-
           \end{pmatrix} \,,
&
 \Omega &= \begin{pmatrix}
      1 - \Lambda^+ & \;\, -{\mathrm{i}}\Lambda^- \\ -{\mathrm{i}}\Lambda^- & 1 + \Lambda^+
     \end{pmatrix} \,,
\end{align}
The result of the integral in Eq.~\eqref{eq:O:4} is $ \sqrt{\frac{\pi^{4N}}{\det \Omega}} e^{\frac{1}{4} \vek w^T \Omega^{-1} \vek w} $
and Eq.~\eqref{eq:O:3} can be cast in the form 
\begin{equation}
  \label{eq:overlap:groundstate:result1}
  \mathcal O(t) = \frac{Z^2}{\sqrt{\det \Omega}} e^{2{\rm Re}\{G_0\}} e^{\frac{1}{4} \vek w^T \Omega^{-1} \vek w} \,,
\end{equation}
with 
 \begin{equation*}
G_0 =  \sum_{jk} \frac{A_{jk}}{2} \beta^e_j{}^*\beta^e_k{}^* - \sum_j \frac{|\beta^e_j|^2}{2} \,.
\end{equation*}
Using Eq.\eqref{eq:groundstateoverlap} in Eq.~\eqref{eq:overlap:groundstate:result1}, the visibility can then be cast in the form of Eq.~\eqref{eq:overlap:groundstate:result2}.

The convergence of the integral in Eq.~\eqref{eq:O:4} is verified by showing that the matrix $\Omega = 1 - \mathrm{B}$, with 
\begin{equation}
 \mathrm{B} = \begin{pmatrix}
       \Lambda^+ & \;\, {\mathrm{i}}\Lambda^- \\ {\mathrm{i}}\Lambda^- & - \Lambda^+
     \end{pmatrix} \,,
\end{equation}
has only eigenvalues whose real parts are greater than zero. For this purpose we consider the spectral radius of $\mathrm{B}$, $\rho(\mathrm{B}) = \max \{ \modulus{\lambda_{\mathrm{B}}} \} $, where $ \lambda_{\mathrm{B}}$ is an eigenvalue of $\mathrm{B}$ and  which fulfills $ \rho(\mathrm{B}) \leq \matrixNorm{\mathrm{B}}$ for any matrix norm~\cite{HornJohnson}. 
If $ \matrixNorm{\mathrm{B}} < 1 $, it follows that all eigenvalues of $\mathrm B$ lie within a circle with radius $\rho(\mathrm{B})<1$ centered around 1. Then, all the real parts of all eigenvalues of $ \Omega = \bigl ( 1 - \mathrm{B} \bigr ) $ are greater than zero.
$\mathrm{B}$ can be brought to block-diagonal form $\mathrm{D}_{\mathrm{B}}$ by a similarity transformation with an orthogonal matrix~$\mathrm{M}_{\mathrm{B}}$:
\begin{equation}
 \mathrm{B} = \frac{1}{2}\begin{pmatrix}
      1 & \phantom{-}1 \\ {\mathrm{i}} & -{\mathrm{i}}
     \end{pmatrix}\begin{pmatrix}
      \mathrm{A}(t) & 0 \\ 0 & \mathrm{A}
     \end{pmatrix}\begin{pmatrix}
      1 & \phantom{-}{\mathrm{i}} \\ 1 &-{\mathrm{i}}
     \end{pmatrix} \,.
\end{equation}
\vspace{1pt}\\
Thus $\matrixNorm{\mathrm{B}} = \matrixNorm{\mathrm{M}_{\mathrm{B}} \mathrm{D}_{\mathrm{B}} \mathrm{M}_{\mathrm{B}}^T} \leq \matrixNorm{\mathrm{M}_{\mathrm{B}}}\matrixNorm{\mathrm{D}_{\mathrm{B}}}\matrixNorm{\mathrm{M}_{\mathrm{B}}}$ by the submultiplicativity of the matrix norm. The spectral norm of the orthogonal matrices is 1, and the spectral norm of $\mathrm{D}_{\mathrm{B}}$, $\matrixNorm{\mathrm{D}_{\mathrm{B}}}_2 = \max \{\matrixNorm{\mathrm{A}(t)}_2,\matrixNorm{\mathrm{A}}_2\}$, but since $\matrixNorm{\mathrm{A}(t)}_2=\matrixNorm{\mathrm{A}}_2$, we have $\matrixNorm{\mathrm{B}}_2 = \matrixNorm{\mathrm{A}}_2$.

We now proceed to perform a Taylor expansion of Eq.~\eqref{eq:overlap:groundstate:result2} for short times. For this purpose  we first bring expression~\eqref{eq:overlap:groundstate:result2} into a more convenient form, using
the definitions
\begin{align}
  \Xi &= 1 + \Lambda^+  \,, &
  \Upsilon &= 1 - \Lambda^+ \,.
\end{align}
The determinant and the inverse matrix can be calculated with the help of the corresponding identities for a partitioned matrix~\cite{HendersonSearleSIAM1981}, 
\begin{equation}
 \det \Omega = \det \Xi \cdot \det \Theta \,,
\end{equation}
and
\begin{equation*}
 \Omega^{-1} = 
    \begin{pmatrix}
      \Theta^{-1} & {\mathrm{i}} \Theta^{-1}\Lambda^-\Xi^{-1} \\
      {\mathrm{i}} \Xi^{-1}\Lambda^-\Theta^{-1} & \quad \Xi^{-1} - \Xi^{-1}\Lambda^- \Theta^{-1} \Lambda^- \Xi^{-1}
    \end{pmatrix} \,,
\end{equation*}
where $\Theta$ is the Schur complement of $\Xi$ given by 
\begin{equation}
 \Theta = \Upsilon +  \Lambda^- \Xi^{-1} \Lambda^- \,.
\end{equation}
The equations hold provided that $\Xi$ and $\Theta$ are non-singular, which is true as shown in Appendix~\ref{sec:disentangling}.

Expanding the overlap 
 around $t=0$, we find 
\begin{equation}
 \mathcal O(t) \approx 1 - {\mathrm{i}} \mathcal O_1 t - \frac{1}{2} \mathcal O_2 t^2 \,,
\end{equation}
with $\mathcal O_1 = \dot{\mathcal O}(0)$ and $\mathcal O_2 = \ddot{\mathcal O}(0)$, which leads to the expression of the visibility in Eq.~\eqref{eq:V:tshort}, where $v=-(\mathcal O_2 - \mathcal O_1^2) $.


\end{document}